\documentclass[fleqn,11pt]{wlscirep}
\hypersetup{final}
\usepackage[utf8]{inputenc}
\usepackage[T1]{fontenc}
\usepackage{bm}
\usepackage{caption}
\usepackage{subcaption}
\usepackage{graphicx}
\usepackage{amsmath}
\usepackage{ulem}

\usepackage[dvipsnames]{xcolor}
\usepackage{gensymb}
\usepackage{color,soul}
\newcommand{\equalcontrib}{\textsuperscript{\#}}
\title{Piezoelectric truss metamaterials: data-driven design and additive manufacturing} 

\DeclareMathOperator*{\argmaxA}{arg\,max}
\author[1,2,3]{Saurav Sharma\equalcontrib}
\author[1]{Satya K. Ammu\equalcontrib}
\author[2]{Prakash Thakolkaran}
\author[*,3]{Jovana Jovanova}
\author[*,1]{Kunal Masania}
\author[*,2]{Siddhant Kumar}
\affil[1]{Shaping Matter Lab, Faculty of Aerospace Engineering,
Delft University of Technology, 2629 HS, Delft, The Netherlands.}
\affil[2]{Department of Materials Science and Engineering, Faculty of Mechanical Engineering, Delft University of Technology, 2628 CD, Delft, The Netherlands.}
\affil[3]{Department of Maritime and Transport Technology, Faculty of Mechanical Engineering, Delft University of Technology, 2628 CD, Delft, The Netherlands.}

\affil[*]{J.Jovanova@tudelft.nl, K.Masania@tudelft.nl, Sid.Kumar@tudelft.nl \newline \equalcontrib These authors contributed equally.}

\newcommand{\rev}[1]{{{#1}}}

\keywords{Piezoelectric metamaterials, Inverse design, In-gel 3D printing, Machine learning}

\newcommand{\boldface}[1]{\boldsymbol{#1}}  %

\newcommand{\bfe}{\boldface{e}}

\newcommand{\bfp}{\boldface{p}}

\newcommand{\bfx}{\boldface{x}}

\newcommand{\bfz}{\boldface{z}}
\newcommand{\bfA}{\boldface{A}}

\newcommand{\bfR}{\boldface{R}}

\newcommand{\bfU}{\boldface{U}}
\newcommand{\bfV}{\boldface{V}}

\newcommand{\bfXi}{\boldsymbol{\Xi}}

\newcommand{\calF}{\mathcal{F}}

\newcommand{\calP}{\mathcal{P}}

\newcommand{\dsK}{\mathbb{K}}

\newcommand{\dsV}{\mathbb{V}}

\newcommand{\Rset}{\mathbb{R}}

\newlength{\boxwidth}
\def\btheorem{\begin{theorem}}
\def\etheorem{\end{theorem}}
\def\blemma{\begin{lemma}}
\def\elemma{\end{lemma}}
\def\bproposition{\begin{proposition}}
\def\eproposition{\end{proposition}}
\def\bcorollary{\begin{corollary}}
\def\ecorollary{\end{corollary}}
\def\bdefinition{\begin{definition}}
\def\edefinition{\end{definition}}
\def\bexample{\begin{example}}
\def\eexample{\end{example}}
\def\bremark{\begin{remark}}
\def\eremark{\end{remark}}

\DeclareMathOperator{\diag}{diag}

\newcommand{\be}{\begin{equation}}
\newcommand{\ee}{\end{equation}}
\newcommand{\beq}{\begin{eqnarray}}
\newcommand{\eeq}{\end{eqnarray}}
\newcommand{\bem}{\begin{multline}}
\newcommand{\eem}{\end{multline}}
\newcommand{\ba}{\begin{align}}
\newcommand{\ea}{\end{align}}

\begin{abstract}
\rev{The inherent directionality of piezoelectric materials is constrained by the symmetry of their crystal structure, which limits the property space in natural piezoelectric materials. To alleviate this limitation, one could leverage geometry or architecture at the mesoscale. Here, we present a framework for designing and 3D-printing piezoelectric truss metamaterials with customizable anisotropic responses. We employ generative machine learning to design truss metamaterials and achieve unconventional behaviors, including auxetic, unidirectional, and omnidirectional piezoelectricity. Then, we develop an in-gel-3D printing method to fabricate these structures using a composite slurry of photo-curable resin and lead-free piezoelectric particles. We achieve an improvement of over 48\% in the specific hydrostatic piezoelectric coefficient in optimized metamaterials over bulk lead zirconate titanate (PZT), and the rare phenomenon of higher transverse piezoelectric coefficients than the longitudinal coefficient. Our approach enables customizable piezoelectric responses and paves the way towards the development of a new generation of electro-active animate materials.}

\textbf{Keywords:} Piezoelectricity, Metamaterials, Inverse design, In-gel 3D printing, Machine learning. 
\end{abstract}

\begin{document}

\flushbottom
\maketitle

\section*{Introduction}
Piezoelectric materials demonstrate a separation of positive and negative charge centers at the atomic scale in response to mechanical deformation, which facilitates the two-way coupling between mechanical and electrical energies. This makes them an indispensable component in a wide range of engineering applications across domains, such as biomedical \cite{kim2020biomolecular, tandon2018piezoelectric, xu2021construction}, aerospace \cite{boukabache2014toward, elahi2018response}, electronics \cite{jackson2013flexible, kim2012piezoelectric}, building structures \cite{kim2018optimized, yang2018preliminary}, and electro-chemistry \cite{qian2020piezoelectric}. The search for superior piezoelectric properties has been a topic of high interest among the scientific community. Yet, the progress has been primarily restricted to improving the performance of existing piezoelectric materials by either material synthesis-based methods \cite{hu2019improved, shi2021interface, abbasipour2019improving, krishnaswamy2019improving, chen2018y2o3} or macro-scale design and optimization methods \cite{wang2018optimization, de2019topology, yoon2018multiphysics}. With the advent of technologies such as micro and nanoelectromechanical systems (MEMS and NEMS) and micro-robotics, the need for exotic and tunable electromechanically responsive systems is pushing the limits of functionalities beyond what current materials can offer. The primary restricting factor for the range of piezoelectric response in conventional piezoelectric materials is their crystal symmetry, which dictates the directionality of electromechanical coupling.

The piezoelectric coupling correlates six mechanical stresses or strains to the three electric field components. This correlation is defined by the piezoelectric tensor, which has 18 unique coefficients correlating the electrical and mechanical state variables. Commonly used piezoelectric materials such as dielectric ceramics, e.g., lead zirconate titanate (PZT), BaTiO\textsubscript{3} (BT), PMN-PT, and polymers, e.g., polyvinylidene fluoride (PVDF) have only five non-zero piezoelectric coefficients, due to their crystal symmetry. These coefficients have fixed signs (positive or negative) which cannot readily be flipped. This makes certain desirable functionalities in piezoelectric materials rare. For example, the phenomenon of negative piezoelectricity, capable of generating compression in response to a positive electric field, has only been observed in PVDF and its co-polymers \cite{katsouras2016negative}. In addition to the individual coefficients, several of their desirable combinations are unavailable in conventional materials, such as observing piezoelectric response only in a particular direction and eliminating the noise signal from the other directions and zero longitudinal piezoelectricity to circumvent microphony \cite{xu20163d, chong2002pyroelectric}. Thus, realizing the full anisotropic design space of piezoelectric materials could be beneficial for several applications and warrants the development of an effective design and fabrication route.

While changing or designing the arrangement of ions in the material crystals may not be physically feasible to achieve tailorable properties, it is possible to design the architecture of the material at a mesoscale. In this context, metamaterials, also known as architected materials, emerge as a promising way to design for tailored piezoelectric properties. Metamaterials are constructed by tessellations of periodic or aperiodic unit cells large enough to be manufacturable yet small enough to constitute a pseudo-homogeneous response at the macro scale. One notable class of these metamaterials is beam-based truss metamaterials (constructed by a spatial arrangement of connected struts), which have been shown to be extremely light-weight and mechanically strong \cite{bastek2022inverting, zheng2023unifying, shaikeea2022toughness}.

The architecture of a truss metamaterial is defined by the combination of the geometry and topology of the unit cells, where the geometry means the position of the junctions or nodes of the struts, and topology is the connectivity of these nodes. Truss metamaterials have been vastly explored for exotic and tunable mechanical properties \cite{zheng2023unifying,shaikeea2022toughness, bertoldi2017flexible, zheng2014ultralight, rafsanjani2015snapping, rocklin2017transformable, dudek2022micro}. The development of piezoelectric metamaterials requires special considerations in design and fabrication due to multiphysics coupling and the brittle nature of piezoelectric materials. Cui et al. have demonstrated the tailorable response of architected piezoelectric metamaterials and their applications in designing robotic metamaterials \cite{cui2019three, cui2022design}. While the presented concepts are promising, the design process has remained either ad-hoc or based on a simple parametrization of a small design space. The main challenge is that of inverse design, namely,  exploration of the infinite number of possible designs to discover the ones with desirable properties.

The inverse design problem has been tackled in mechanical metamaterials \cite{zheng2023unifying, bastek2022inverting}, however for piezoelectric materials, it becomes more complicated due to multiphysics coupling and poling orientation dependence of piezoelectric properties. To solve this challenge, we propose a generative machine learning (ML) approach for designing the piezoelectric response in architected materials. We use two distinct descriptions of the truss metamaterials, where one provides diverse piezoelectric responses while the other facilitates a broad range of properties and better manufacturability. The design-property pairs of these truss metamaterials are used to train the ML framework, which can generate designs corresponding to desired properties. Having established the design framework, we further address the next major challenge in the development of piezoelectric metamaterials, which is their fabrication.

Additive manufacturing techniques, such as stereolithography (SLA), digital light processing (DLP), and fused filament deposition (FFF), offer great freedom in fabricating a broad range of topologies. Doing so with piezoelectric materials is challenging for several reasons, such as the requirement of high-temperature processing in ceramics and lower piezoelectricity in polymers, low mechanical integrity due to their brittleness, and the requirement of poling post-manufacturing. Printing polymer-ceramic piezoelectric composites with a photo-curable resin ink infused with piezoelectric particles has been shown recently for manufacturing spinodoid topologies with bi-continuous features \cite{shi20243d}. This approach requires high precision to manufacture slender structural members of low relative density metamaterials, such as those based on truss architectures. Moreover, their printing requires recoating of the printing ink for each layer due to the necessary high particle loading, which leads to excess material usage and a cost-intensive setup \cite{cui2019three, shi20243d}. Another difficulty in achieving continuous SLA/DLP printing with piezoelectric inks arises from the contrast in optical and rheological properties of the piezoelectric particles and photo-curable resins. Direct ink writing (DIW) offers a better approach for this purpose, where composite inks can be extruded to print shapes in 3D space \cite{ammu20243d}. This approach has been shown to be effective by Tao et al. \cite{tao2022multi} for printing piezoelectric materials with relatively simpler shapes. However, sustaining the accurate shape and ensuring connectivity becomes challenging when the whole structure cannot be printed continuously, such as the truss structures we propose. 

To achieve this, we develop an in-gel 3D printing technique to enable the additive manufacturing of piezoelectric truss metamaterials. A photosensitive resin-based ink infused with piezoelectric microparticles is synthesized and used to extrude the desired shapes. With an optimized print path and rheology, the printing technique enables connectivity of the struts while the shape accuracy is preserved due to structural support provided by the gel. Our inverse design framework, combined with the developed 3D printing methodology, could successfully design and fabricate unique piezoelectric metamaterials with tunable piezoelectric response. For example, we designed metamaterial unit cells with maximized hydrostatic piezoelectric coefficient, transverse piezoelectric coefficient higher than longitudinal, auxetic piezoelectric response, and selective piezoelectric coefficients.

\section*{Results}

\subsection*{Piezoelectric truss metamaterials}
\begin{figure}
    \centering
    \includegraphics[width=\linewidth]{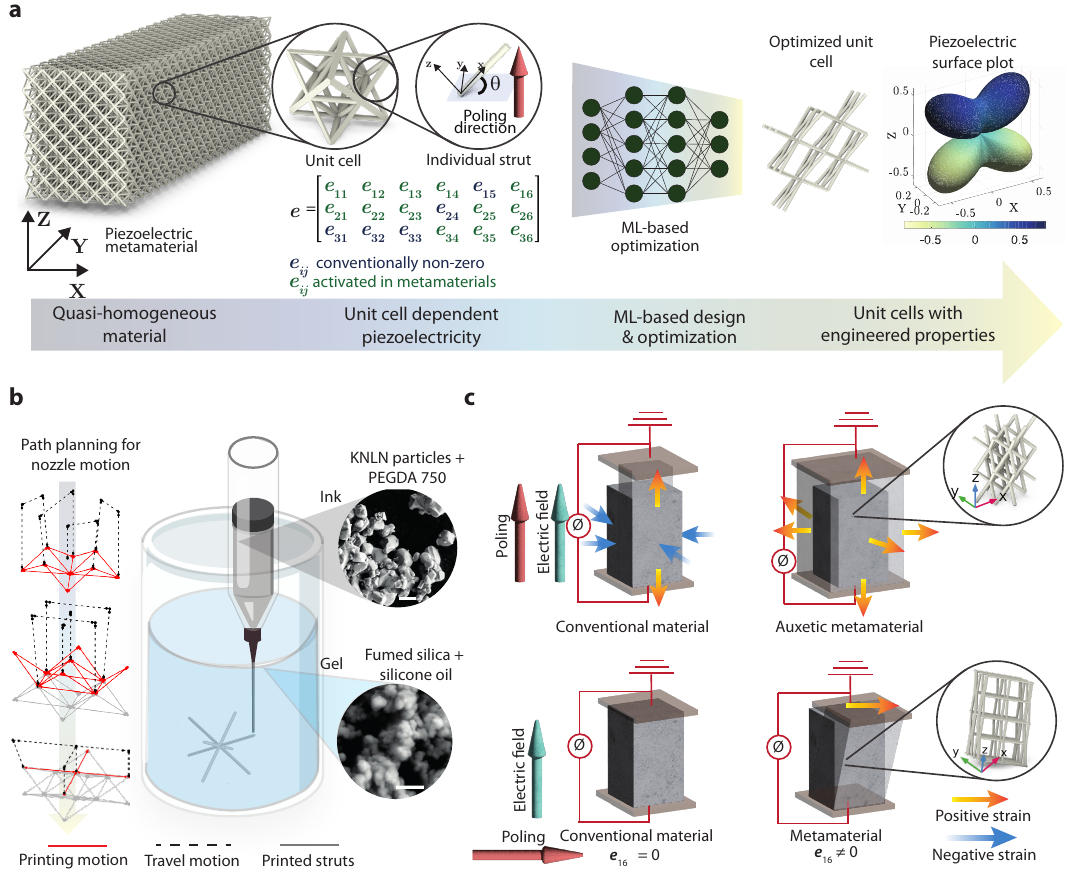}
    \captionsetup{justification=justified}
    \caption{\textbf{Concept of piezoelectric truss metamaterials for tailoring the electromechanical behavior.} \textbf{a.} Illustration of a piezoelectric metamaterial made by tessellating octet unit cells in three orthogonal directions. The different levels of structural features are shown in the insets. These mesoscale features can be tuned to selectively activate different components of the piezoelectric property matrix, in contrast to the only five non-zero coefficients in most of the common piezoelectric materials. ML is then used to design unit cells with desired properties, represented by the piezoelectric surface plot on the right. \textbf{b.} In-gel direct-ink-writing-based 3D printing of the designed metamaterials is performed by extruding the piezoelectric ink in a support gel by following the planned path for minimal travel motion. The piezoelectric composite ink is made of KNLN particles and PEGDA 700 matrix, as shown in the inset (scale bar 2 $\mu$m). The ink is then extruded in a support gel made of fumed silica and silicon oil (scale bar 200 nm). \textbf{c.} Examples of piezoelectric metamaterials made by the truss unit cells exhibiting behaviors not readily available in conventional materials. The first example (top) shows the auxetic behavior of a truss metamaterial. While conventional materials, under an applied electric field in the poling direction, expand in the direction of the electric field and contract along the two normal directions, the metamaterial made of the auxetic unit cell expands in all three directions. The second example (bottom) shows a metamaterial with a non-zero $e_{16}$ coefficient, exhibiting shear deformation in response to an electric field applied in direction 1.}
    \label{fig:intro}
\end{figure}

The homogenized properties of truss metamaterials at the macro scale depend on the topology and the geometry of the unit cells. In the case of piezoelectric metamaterials, the orientation dependence of the piezoelectric property tensor of individual struts also becomes important. \figureautorefname~\ref{fig:intro}\textbf{a} demonstrates the concept of piezoelectric truss metamaterials. The inclination of the individual struts activates multiple modes of piezoelectricity simultaneously, as demonstrated in \figureautorefname~\ref{fig:intro}\textbf{a}. We define a global coordinate system (X, Y, Z) and a local coordinate system for individual struts (x, y, z) where x-axis aligns with the length of the strut (Supporting Information Figure S1). The two-level orientation dependence of the piezoelectric tensor arises from the fact that the piezoelectric tensor is defined with reference to the poling direction, which is parallel to the global Z axis. As the orientation of the individual struts with respect to the poling direction can be different, the piezoelectric coefficients can have different values for individual struts, which contributes to the complexity of the design-property relationship.

The two most common forms of constitutive equations of piezoelectricity are strain-charge and stress-charge forms, which can be written in Einstein summation notation as
\begin{subequations}
\label{eq:strain-charge}
    \begin{align}
    \epsilon_{ij} &= S_{ijkl}^E \sigma_{kl} + d_{ijk} E_k, \tag{1a} \\
    D_i &= d_{ijk} \sigma_{jk} + \kappa_{ik}^{\sigma} E_k, \tag{1b}
    \end{align}
\end{subequations}
and,
\begin{subequations}
    \label{eq:stress-charge}
    \begin{align}
    \sigma&_{ij}=C_{ijkl}^E\epsilon_{kl}-e_{ijk}E_k, \tag{2a} \\
    D&_i=e_{ijk}\epsilon_{jk}+\kappa_{ik}^{\epsilon}E_k. \tag{2b} 
    \end{align}
\end{subequations}
Here $\epsilon_{ij}$ and $\sigma_{ij}$ denote the mechanical strain and stress tensors, and the electric field and electric displacement vectors are designated as $E_k$ and $D_k$. $C_{ijkl}$, $e_{ijk}$, $d_{ijk}$ and $\kappa_{ik}$ denote mechanical stiffness, piezoelectric coupling coefficients in stress charge and strain charge forms, and dielectric permittivity of the material. The indices $i$, $j$, $k$, $l$ $\in \{1,2,3\}$, while the superscripts $(\cdot)^E$, $(\cdot)^\sigma$, and $(\cdot)^\epsilon$ indicate the properties measured at a constant electric field, stress, and strain, which are omitted hereafter for the sake of brevity. To concur with the finite element framework used for homogenization, the stress-charge form of piezoelectricity in Eq. \ref{eq:stress-charge} is used. Theoretically, it has been shown that the piezoelectric tensor can be tuned by controlling the direction of poling of the
material \cite{sharma_design_2021}. We apply poling along the global $Z$ axis, so the effective piezoelectric coefficient in the local (x,y,z) coordinate system of a beam, denoted by $e'$ can be computed as
\be
\label{eq:e_dash}
    e'_{pqr}=R_{pi}R_{qj}R_{rk}e_{ijk}^{\text{base}}.
\ee
Here, $e_{ijk}^{\text{base}}$ are the components of the piezoelectric tensor of the base material, $\bfR\in\text{SO(3)}$ is the rotation matrix (see Supporting Information Eq. 2) containing the direction cosines of the local coordinate axes (x, y, z), with respect
to global coordinate axes (X, Y, Z). Exploiting the symmetry of stress and strain components (e.g., $\sigma_{ij}=\sigma_{ji}$), Eq. \ref{eq:stress-charge} can be simplified and written in matrix form as

\be
\label{eq:Main_stress-charge}
\begin{bmatrix}
    \sigma_{11} \\
\sigma_{22} \\
\sigma_{33} \\
\sigma_{23} \\
\sigma_{31} \\
\sigma_{12} \\
D_{1} \\
D_{2} \\
D_{3} \\
\end{bmatrix}
=
\begin{bmatrix}
C_{11} & C_{12} & C_{13} & C_{14} & C_{15} & C_{16} & -e_{11} & -e_{21} & -e_{31} \\
C_{21} & C_{22} & C_{23} & C_{24} & C_{25} & C_{26} & -e_{12} & -e_{22} & -e_{32} \\
C_{31} & C_{32} & C_{33} & C_{34} & C_{35} & C_{36} & -e_{13} & -e_{23} & -e_{33} \\
C_{41} & C_{42} & C_{43} & C_{44} & C_{45} & C_{46} & -e_{14} & -e_{24} & -e_{34} \\
C_{51} & C_{52} & C_{53} & C_{54} & C_{55} & C_{56} & -e_{15} & -e_{25} & -e_{35} \\
C_{61} & C_{62} & C_{63} & C_{64} & C_{65} & C_{66} & -e_{16} & -e_{26} & -e_{36} \\
e_{11} & e_{12} & e_{13} & e_{14} & e_{15} & e_{16}  & \kappa_{11} & \kappa_{12} & \kappa_{13} \\
e_{21} & e_{22} & e_{23} & e_{24} & e_{25} & e_{26}  & \kappa_{21} & \kappa_{22} & \kappa_{23} \\
e_{31} & e_{32} & e_{33} & e_{34} & e_{35} & e_{36}  & \kappa_{31} & \kappa_{32} & \kappa_{33} \\
\end{bmatrix}
\begin{bmatrix}
\varepsilon_{11} \\
\varepsilon_{22} \\
\varepsilon_{33} \\
2\varepsilon_{23} \\
2\varepsilon_{31} \\
2\varepsilon_{12} \\
E_{1} \\
E_{2} \\
E_{3} \\
\end{bmatrix}.
\ee
In the above equation, the 18 unique coefficients of the piezoelectric tensor are written in the form of a \rev{$3\times 6$} piezoelectric matrix, which will be used hereafter to denote the piezoelectric property space. The conversion of piezoelectric coefficients from stress-charge to strain-charge and vice versa are shown in Supporting Information Eq. (5). The effective electrical, mechanical, and electromechanical properties of a unit cell defined by its geometry and topology
are computed using a finite element-based homogenization with periodic boundary conditions, as explained in the Supporting Information Section 1. However, a method to discover the geometry and topology of the unit cells corresponding to desired properties is needed to achieve a tailored piezoelectric response. Next, we develop an ML-based design methodology to achieve this.

\subsection*{ML-based optimization for tailored piezoelectricity} 
Truss metamaterials offer great tunability of physical properties due to virtually infinite possible topologies and geometries. This comes at the cost of higher complexity in the design space and makes design and optimization difficult. The parameterization of these truss metamaterials is often discrete and high-dimensional. The complexity is further enhanced by the thin features, which makes traditional optimization methods computationally expensive to implement for inverse designing to achieve desired properties.
We explore the vast and highly complex design space of truss metamaterials and their respective piezoelectric properties by selecting two fundamentally different design spaces, each with its own advantages. These design spaces were originally developed for mechanical metamaterials and were used to train inverse design models for designing elastic properties \cite{bastek2022inverting, zheng2023unifying}. We adapt these design spaces and develop the optimization framework for designing the piezoelectric behavior of truss metamaterials. The first design space, based on geometric transformations, provides a broader range of piezoelectric responses and can activate all 18 piezoelectric coefficients. However, these geometric transformations lead to aspect ratios and rotations that are unfavorable for manufacturing. The second design space is restricted to a cubic symmetry and thus can only activate the five already non-zero coefficients of the base materials, with a broader range of values and improved manufacturability compared to the first design space. In the following, we present two different ML frameworks to suit the nature of the parameterizations and efficiently optimize in these design spaces.

\textbf{Design space I:} The first design space, introduced by Bastek et al.\cite{bastek2022inverting}, which was originally inspired by Zok et al.\cite{zok2016periodic}, is based on the superposition of a small set of elementary unit cells, which are subsequently stretched and rotated twice to get the final unit cell. Each unit cell is described by the following design parameters: \textit{(i)} three elementary unit cells {$L_1,L_2,L_3\in\{1,\dots,7\}$} {chosen out of 7 options},
    \textit{(ii)} their respective tessellations  {$t_1,t_2,t_3\in\{1,2\}$} {(i.e., each cell tessellated as $1\times1\times1$ or $2\times2\times2$)} following which they are superposed,
    \textit{(iii)} the eigenvalues $U_1,U_2,U_3>0$ of the first affine stretch tensor $\bfU=\diag(U_1,U_2,U_3)$,
    \textit{(iv)} the first rotation tensor $\bfR_I\in \text{SO}(3)$,
    \textit{(v)} the eigenvalues $V_1,V_2,V_3>0$ of the second affine stretch tensor $\bfV=\diag(V_1,V_2,V_3)$,
    \textit{(vi)} the second rotation tensor $\bfR_{II}\in \text{SO}(3)$, and lastly,
    \textit{(vii)} the relative density $\rho$ of the final unit cell  achieved by choosing uniform diameter struts. The rotation tensors are defined by three unique parameters in axis-angle representation. 
    Thus each truss lattice is defined by the set of design parameters $\Theta=\{L_1,L_2,L_3,t_1,t_2,t_3,U_1,U_2,U_3,\bfR_I,V_1,V_2,V_3,\bfR_{II},\rho\}$.
     Three example unit cells generated using this parametrization and their corresponding parameters are shown in Figure S3\textbf{a}. Using this approach, one can obtain 262 unique topologies and virtually infinite designs. We used a set of $3,000,000$ unit cell designs generated with this parametrization and performed numerical homogenization to obtain the pairs of unit cell designs and piezoelectric properties. Due to the rotation and stretching, we observe that all 18 components of the effective piezoelectric matrix $\bfe$ are activated. Further, we note that certain components correlate highly with others (see Section 3 of the Supporting Information).
     
We train a forward model $\calF_\omega:\Theta\rightarrow \bfe$ based on a neural network (NN)  that maps both geometric and topological design parameters $\Theta$ of the truss lattices to the effective piezoelectric matrix $\bfe$. Here, $\omega$ denotes the trainable parameters (weights and biases) of the NN architecture.  
We use a multi-layer perceptron (MLP) architecture for the NN; see Supporting Information Section 2 for details. In Supporting Information Figure S2\textbf{a}, we illustrate the prediction performance of the forward model with parity plots.

\begin{figure}
\vspace{-16pt}
    \centering
    \includegraphics[width=\linewidth]{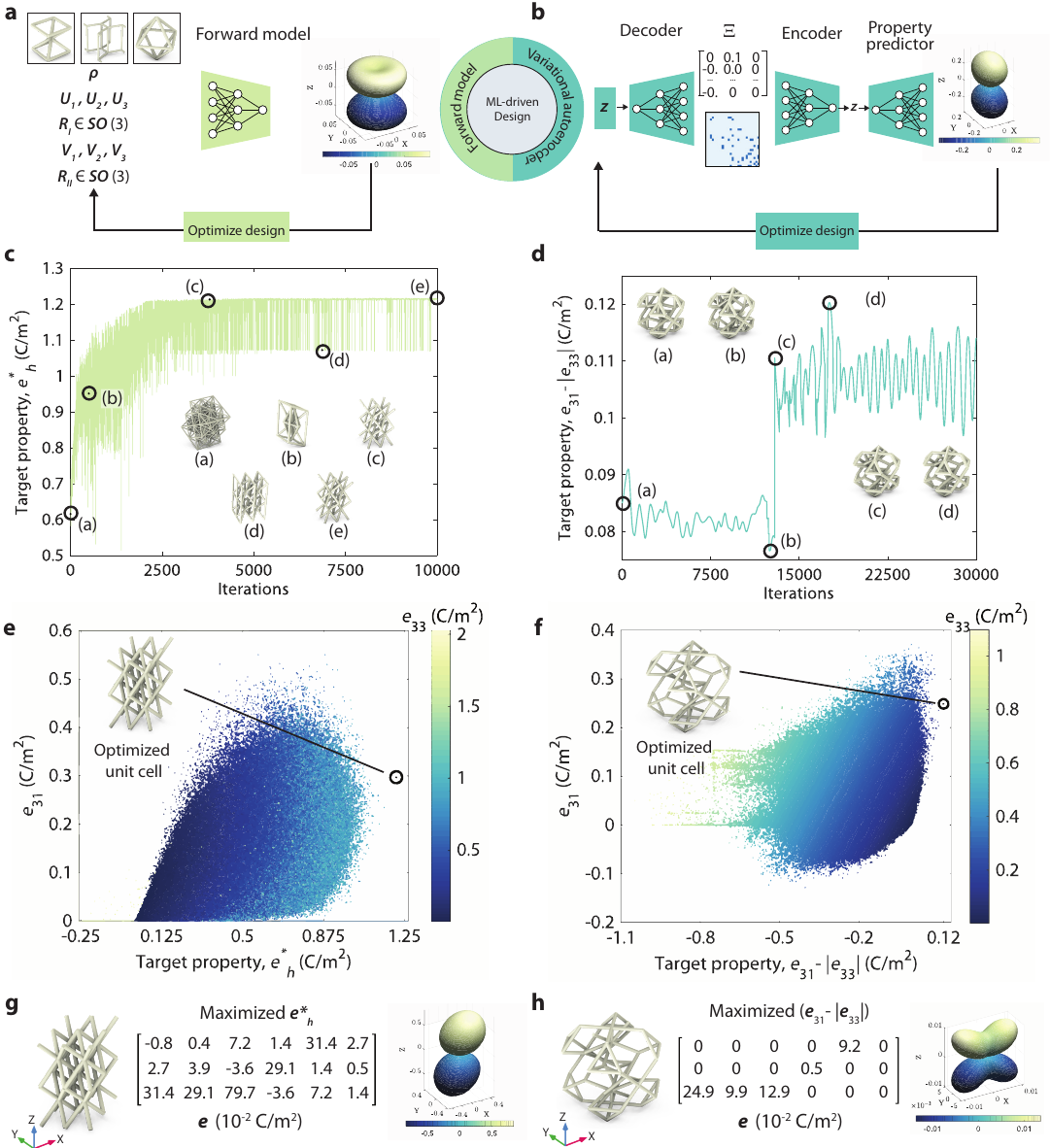}
    \captionsetup{justification=justified}
    \caption{\textbf{ML-based optimization of piezoelectric metamaterials}. \textbf{a.} The optimization framework employed for the design space I. A pre-trained forward neural network (NN) is used to perform gradient-based optimization. \textbf{b.} The variational autoencoder (VAE) architecture employed to perform optimization in the lower-dimensional and continuous latent space for design space II. The additional encoding-decoding step is performed to ensure the validity of the predicted designs. The trace followed during the optimization, and different designs traversed during the path for \textbf{c.} Dataset I and \textbf{d.} Dataset II. \textbf{e.} The $e_{31}$ vs.~target property $e_h^*$ for dataset I, and \textbf{f.} the $e_{31}$ vs.~target property ($e_{31}-|e_{33}|$) landscape for dataset II. In both cases, the achieved optimized lattices, marked by the black circles, lie beyond the dataset and demonstrate the extrapolatory capability of the optimization framework. \textbf{g.} An example optimized unit cell is shown, where the hydrostatic piezoelectric coefficient $e_h^*$ is maximized. \textbf{h.} An example optimized lattice with the optimization objective to maximize ($e_{31}-|e_{33}|$).}
    \label{fig:ML}
\end{figure}

Once trained, the forward model can be used as a surrogate structure-property map for design optimizations, as shown in \figureautorefname~\ref{fig:ML}\textbf{a}. Here, we demonstrate this capability by designing a truss metamaterial with maximal hydrostatic coefficient $e_h$,
\be
e_h[\bfe] = e_{31}+e_{32}+e_{33},
\ee
which quantifies the energy harvesting performance of a piezoelectric material under hydrostatic pressure, with higher being better. For converse piezoelectricity (actuation), $e_h$ signifies the volumetric expansion under an electric field in direction 3. Since in most piezoelectric materials, $e_{33}$ is positive while $e_{31}$ and $e_{32}$ are negative, very low magnitudes of $e_h$ are observed, leading to poor performance. To achieve an improved $e_h$, we formalize the design optimization problem as:

\begin{align}
\begin{split}
\label{eq:ds1_optim}
\Theta^* \leftarrow & \argmaxA_\Theta\ e_h^*[\bfe], \qquad \text{with}\quad \bfe = \calF_\omega(\Theta) \quad \text{and} \\
& \  e_h^*[\bfe] = e_h[\bfe] - \lambda_1 \underbrace{\left( (e_{31} - e_{32})^2 + (e_{31} - e_{33})^2 + (e_{32} - e_{33})^2)\right)}_{\text{magnitude regularization}} - \lambda_2 \underbrace{\left(\sum_{i=1}^{6} e_{1i}^2 + \sum_{i=1}^{6} e_{2i}^2 + \sum_{i=4}^{6} e_{3i}^2\right)}_{\text{shrinkage regularization}},
\end{split}
\end{align}

where $\Theta^*$ is the set of optimized design parameters. The second term with scaling hyperparameter $\lambda_1>0$ aids in regularizing the magnitudes of all three components so that we can achieve close-to-uniform volumetric expansion. The third term with scaling hyperparameter $\lambda_2>0$ serves as a shrinkage regularization to prevent distortions such as shear during actuation. 
To avoid extreme aspect ratios and improve the manufacturability of the optimized designs, we constrained the optimization process such that the stretch values $\bfU$ and $\bfV$ are scaled down by a factor of $\alpha \in (0,1]$. %
The problem in \eqref{eq:ds1_optim} is solved numerically using gradient-based optimization, with Adam optimizer. All hyperparameters, training, and optimization protocols can be found in the Supporting Information Section 2. 

\figureautorefname~\ref{fig:ML}\textbf{c} shows the trace followed by the optimization framework, along with example lattices for different checkpoints. The categorical parameters related to topology lead to fluctuations in the unit cell designs and, consequently, the target property. \figureautorefname~\ref{fig:ML}\textbf{e} shows the distribution of the dataset in the $e_{31}$ vs.~regularized hydrostatic coefficient $e_h^*$ landscape. The optimized lattice lies outside the dataset, demonstrating the extrapolatory capability of the optimization and ML framework. Additional examples of tailored piezoelectricity are presented in the Supporting Information, Section 4, Figure S6, which demonstrate the unconventional behaviors of full anisotropy and unidirectionality of piezoelectricity.

\textbf{Design space II:} The second design space, introduced by Zheng et al.\cite{zheng2023unifying}, is based on a more general approach of generating highly complex unit cells with cubic symmetry.
Each unit cell is created with a fixed set of virtual nodes, which can be offset freely along specific degrees of freedom, and the nodes are activated by connecting them to other nodes. 
Although the generated unit cells are of orthotropic nature, a vast range of stiffness values and orthotropic anisotropy can be achieved. Analogously, the design space is virtually infinite geometrically with at least millions of unique topologies if not more.
Each unit cell is constructed by designing the lattice within the octant of a unit cube, which is then mirrored to obtain the unit cell with cubic symmetry and periodic tileability. 
There are 27 virtual nodes in the octant, i.e., 8 vertex nodes $\{\dsV_1,\dsV_2,\dsV_3,\dsV_4,\dsV_5,\dsV_6,\dsV_7,\dsV_8\}$, 12 edge nodes $\{E_1,E_2,E_3,E_4,E_5,E_6,E_7,E_8,E_9,E_{10},E_{11},E_{12}\}$, 6 face nodes $\{F_1,F_2,F_3,F_4,F_5,F_6\}$, and one body node $\{T_1\}$. While the vertex nodes remain fixed, the edge nodes, face nodes, and body node can move freely on the corresponding edge, face, and inside of the octant, respectively. These node offsets are represented vectorially as $\bfx\in\Rset^{27\times 3}$.
The connectivity is represented by a $27\times 27$ adjacency matrix $\bfA\in\{0,1\}^{27\times 27}$. For $i \neq j$, the entry $A_{ij} = 1$ if nodes $i$ and $j$ are connected by a strut, and $A_{ij} = 0$ otherwise. Diagonal entries follow the rule: $A_{ii} = 1$ if node $i$ is part of the topology, and $A_{ii} = 0$ if it does not exist. A truss lattice is then uniquely defined by the graph $\Xi=(\bfA,\bfx)$. Figure S3\textbf{b} shows three example unit cells generated with this parametrization. Similar to the first design space, we homogenized a set of 965,685 truss metamaterial designs and obtained their piezoelectric properties. 

However, unlike the first design space, the graph-based representations are extremely discontinuous, high-dimensional, and combinatorial. To unify the discrete designs, the variational autoencoder (VAE) framework of Zheng et al.\cite{zheng2023unifying} is employed.  The VAE consists of two neural networks. 
An \textit{encoder} $\mathcal{E}_\phi$, with trainable parameters $\phi$, maps the graph of a truss lattice $\Xi$ to a smooth low-dimensional latent space $\mathbf{z}\in\Rset^d$ of $d$-dimensions, i.e., $\mathbf{z}=\mathcal{E}_\phi(\bfXi)$. A \textit{decoder} $\mathcal{D}_\zeta$ with trainable parameters $\zeta$ reconstructs the graph of the truss lattice from the latent vector, i.e.,  $\Xi=\mathcal{D}_\zeta(\bfz)$. The NNs are trained such that the truss lattices input into the encoder are accurately reconstructed by the decoder, while modeling the latent space to closely follow a standard normal distribution. Given sufficient reconstruction accuracy over a representative dataset, the latent space serves as an informational bottleneck,  providing a continuous and smooth, though abstract, vectorial representation of the design space. New designs can be generated by stochastically sampling the latent space via standard normal distribution and decoding into the graph representation, i.e., $\mathcal{D}_\zeta(\mathbf{z})$ with $\mathbf{z}\sim\mathcal{N}(\mathbf{0},\mathbf{I})$. Simultaneously, a \textit{property-predictor} neural network $\mathcal{P}_\psi:\bfz\rightarrow\bfe$ with trainable parameters $\psi$ is trained to predict the effective piezoelectric matrix $\bfe$ from the latent representation. Figure S2\textbf{b} of the Supporting Information illustrates the prediction performance of the forward model with parity plots. The encoder, decoder, and property predictor are trained jointly, as detailed in Supporting Information Section 2, enabling the latent space to be learned in a manner that facilitates efficient structure-property map exploration.

The pre-trained VAE framework is then used to optimize within the latent space to discover truss lattices that achieve a tailored electromechanical response, as shown in \figureautorefname~\ref{fig:ML}\textbf{b}. The optimization process starts with an initial guess $\mathbf{z}$ from the latent space, which is decoded into the corresponding unit cell architecture $\Xi = \mathcal{D}_\zeta(\mathbf{z})$. To ensure the validity of the architecture\cite{zheng2023unifying}, it is re-encoded back into the latent space $\mathbf{z}' = \mathcal{E}_\phi(\Xi)$.  The property predictor is used to estimate the effective piezoelectric matrix, i.e., $\bfe = \mathcal{P}_\psi(\bfz')$.  The objective function is formulated in terms of $\bfe$. A gradient-based optimization algorithm is then employed to iteratively update the latent vector $\mathbf{z}$; the optimal $\mathbf{z}$ is then decoded into optimal truss lattice design. As an example, we aim to maximize piezoelectric response in transverse mode while minimizing the influence of longitudinal mode, i.e.,
\be
\label{eq:ds2_optim}
\mathbf{z}^* \leftarrow \argmaxA_\mathbf{z} \left({e}_{31} - |{e}_{33}| \right), \qquad \text{with}\quad  \bfe = \calP_\psi(\mathbf{z}') \quad \text{and}\quad \mathbf{z}' = \mathcal{E}_\phi(\mathcal{D}_\zeta(\mathbf{z})).
\ee
Having such a piezoelectric response is advantageous in selectively activating transverse mode while minimizing the influence of longitudinal mode, and thus providing a unidirectional sensor or actuator even under multi-directional loading.

The trace of the optimization in \figureautorefname~\ref{fig:ML}\textbf{d}, along with example lattices from checkpoints along the trace, shows the different designs and their properties traversed by the optimization algorithm in the process. \figureautorefname~\ref{fig:ML}\textbf{f} shows the training dataset properties in the $e_{31}$ vs.~target property $(e_{31}-|e_{33}|)$ landscape. Similar to ~\figureautorefname\ref{fig:ML}\textbf{e}, the optimized unit cell lies outside the bounds of the training dataset, showing an improvement of 18.42\% compared to the best value of target property in the dataset. Details of the optimization and training protocols are presented in Supporting  Information Section 2.

\figureautorefname~\ref{fig:ML}\textbf{g} and \textbf{h} show the optimized unit cells for the representative design problem from both the design spaces, their corresponding piezoelectric matrices, and piezoelectric surface plots. The spatial variation of the surface plots is such that if a vector is drawn from the center of the plot to any point on the surface, the length of the vector represents the magnitude of the longitudinal piezoelectric coefficient, $\Tilde{e}$ in the direction of that vector. Here, $\Tilde{e}(\bfp)$ is defined as the piezoelectric coefficient coupling axial stress in direction $\bfp$ to the electric field in direction $\bfp$. Along the direction $\bfp \in S^2$, where $S^2$ is a unit sphere in three dimensional space, $\Tilde{e} (\bfp)$ is evaluated as
\be
\label{eq:ehat}
\Tilde{e}(\bfp) = \sum_{i,j,k=1}^3 e_{ijk} p_ip_jp_k \ .
\ee

While the first optimization example (\figureautorefname~\ref{fig:ML}\textbf{g}) shows a maximized $e_h$, the second example (\figureautorefname~\ref{fig:ML}\textbf{h}) shows a unidirectionality in electromechanical response, i.e., enhanced $e_{31}$ and minimal $e_{33}$. In the former case, increased $e_h$ and auxetic piezoelectric response provide higher energy harvesting and superior sensing capabilities due to the synergistic effect of transverse and longitudinal modes of operation, which are useful in underwater applications. In the latter case, isolating modes of operation in such a way is useful for transducers for a reduced effect of ambient acoustic and vibration noise, as well as in micro-robotics for specialized motion. Moreover, while $e_{31}$, and $e_{32}$ are negative and $e_{33}$ is positive for most of the conventional piezoelectric materials, in the truss metamaterials presented here, they are all positive, leading to auxetic piezoelectricity. This enables a synergistic interplay of these coefficients in multi-directional loading scenarios, unlike conventional piezoelectric materials, where the interplay is often destructive due to the opposing signs of these coefficients.

To complement the design methodologies and practical realization of truss metamaterials with tailored piezoelectricity, an efficient and accurate fabrication route is needed. Next, we present a custom technique for additively manufacturing these metamaterials.  

\subsection*{In-gel direct ink writing of piezoelectric metamaterials}
Additive manufacturing of piezoelectric truss metamaterials is challenging due to the brittle nature and polarization-dependent behavior of the base piezoelectric materials. Additionally, the 3D slender features with varying orientations complicate the process, especially when relying solely on ink rheology to produce free-standing structures. 

To address this limitation, we developed a direct-ink-writing (DIW) methodology for lead-free piezoelectric transducers. The ink used for printing consists of 3\% Li-substituted sodium potassium niobate K\textsubscript{0.485}Na\textsubscript{0.485}Li\textsubscript{0.03}NbO\textsubscript{3} (KNLN), in combination with UV curable monomer poly-(ethylene glycol) diacrylate, PEGDA-700 (referred to as \textit{piezo ink}). The KNLN particles, with an average size of 3 $\mu$m, were chosen for their high piezoelectric charge constants among lead-free piezo-ceramics and their biocompatibility. Stability and chemical compatibility between the ink and gel are crucial to prevent diffusion or reactions. A hydrophobic medium like silicone oil was chosen to prevent any diffusion of the hydrophilic PEGDA 700 monomer, which was observed in hydrophilic gels. The piezoelectric metamaterials were printed and embedded inside a supporting silicone matrix (referred to as \textit{gel}). This method enables complex geometries without support structures, as struts are printed in a single direction, ensuring directional homogeneity similar to designed metamaterials and fast manufacturing. However, the extruder path must be optimized to avoid collisions and ensure end-connected struts. We used Fleury’s algorithm-based approach proposed by Weeks et al. \cite{weeks2023embedded} to optimize the path traversed by the extruder, avoiding collisions and minimizing travel motions.

The design of the piezoelectric and the support medium must satisfy certain key chemical and rheological requirements to enable the successful embedded printing of lattices. Similar to the requirements of the materials used for DIW in Ammu et al. \cite{ammu20243d}, both the piezo ink and gel need to be shear-thinning yield stress materials, allowing material flow under applied pressure from the extruder. Additionally, the support gel must be thixotropic, enabling it to recover its viscoelastic behavior after shear thinning. This ensures adequate support for the printed ink once the nozzle has moved past, allowing the gel to regain its elasticity and support the weight and shape of the printed ink. The detailed manufacturing steps for the ink and gel are described in the Materials and Methods section.

\begin{figure}
    \centering
    \includegraphics[width=\linewidth]{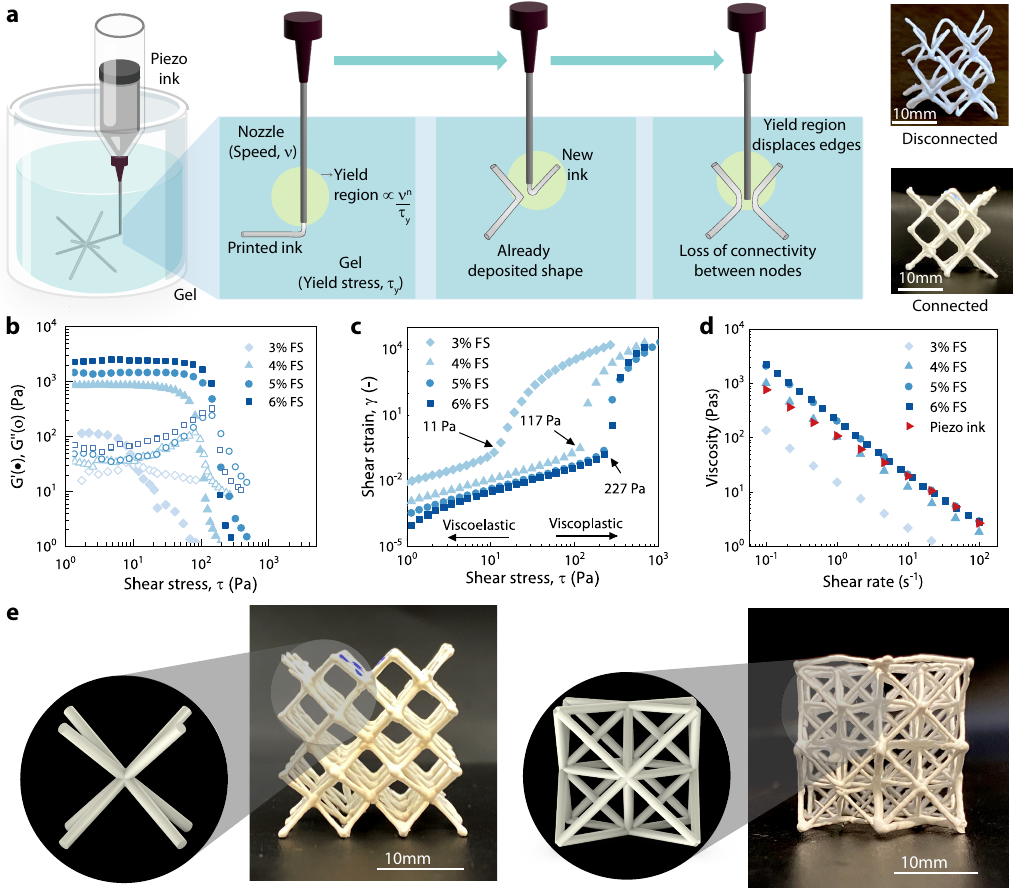}
    \captionsetup{justification=justified}
    \caption{\textbf{Direct ink writing of piezoelectric metamaterials.} \textbf{a.} Schematic of in-gel direct ink writing process, and the optimization of ink rheology and printing parameters to ensure the connectivity at the nodes, \textbf{(b-c).} shear strain variation with applied shear stress for different volume fractions of fumed silica in the gel. \rev{The yield stress points, i.e., 11 Pa, 117 Pa, and 227 Pa, are indicated in \textbf{c}, which are determined from the intersection of tangents at the inflection point of the log–log plots.} \textbf{d.} varying viscosity of the gel with increasing shear stress for different volume fractions of fumed silica. \textbf{e.} \rev{Printed samples of body centered cubic (BCC) (left) and octet (right) lattices. While the tuned rheological properties of the gel allow for the printing of the BCC topology with overhanging struts, the fine-tuned printing speed and ink rheology allow for accurate connectivity of octet topology with a higher number of nodes. The high volume fraction of piezoelectric particles, which leads to the good piezoelectric performance, is detrimental to the printing resolution and can lead to some defects. Nevertheless, we demonstrate the ability to fabricate different topologies with good accuracy and structural integrity.}}
    \label{fig:Printing}
\end{figure}

\subsubsection*{Ink design}
Using the synthesized KNLN particles, we found that ceramic volume fractions in the range between 42 and 47\% exhibit shear thinning behavior along with being able to be printed using our DIW setup. Below this volume fraction, the yield stress of the piezo ink was too low, causing either uncontrolled flow from the nozzle even when at rest or sedimentation of the particles, leading to nozzle clogging and liquid phase migration. The low yield stress of the piezo ink led to fragmentation and beading of the extruded ink due to the yield stress of the ink being lower than the surface tension of the gel, similar to what was observed by Pairam et al. \cite{pairam2013stable}. On the other hand, if the piezo ink's elastic modulus and yield stress are too high, printed filaments get dragged through the matrix by the nozzle, leading to poor connectivity at the nodes as shown schematically in \figureautorefname~\ref{fig:Printing}\textbf{a}. Hence, a volume fraction closer to the lower end of the range, at 43 vol\%, was chosen to maintain a high enough yield stress to allow stable extrusion while not being too viscous. This allows for designing a shear-thinning, thixotropic gel medium whose viscosity can be matched to that of the ink.

\subsubsection*{Gel design}
The gel was designed to have a high shear modulus to sufficiently support the printed filament's weight and prevent sagging, which can occur due to the density mismatch between the ink and the support matrix, especially for ceramics like KNLN with a high density of 4.2 g/cc. Fumed silica (FS) was chosen as the filler for the silicone matrix preparation due to its ability to provide an elastic network structure with sharp shear thinning behavior. Gels with less than 3\% FS showed insufficient yield stress to be viscoelastic, leading to the investigation of gels with more than 3\% FS as support media, as shown in Figure \ref{fig:Printing}\textbf{b}, where the shear storage ($G'$) and loss ($G''$) moduli are plotted against applied shear stress.

The yield stress of the gel must be low enough to allow the nozzle to move through it freely and enable the gel to recover behind the nozzle. Printing inside a lower yield stress gel, such as 3\% FS gel with a yield stress of 11 Pa, resulted in highly distorted and unconnected lattices, as shown in \figureautorefname~\ref{fig:Printing}\textbf{a}. This issue arises because of the large size of the stress field from the moving nozzle in a low-yield stress gel, which interferes with previously extruded ink. The size of the stress field is dependent on the Oldroyd number, defined as \cite{grosskopf2018viscoplastic}
\be
\mathrm{OD} = \frac{\tau_y\Lambda^n}{\dsK U^n} ,
\ee
where $\tau_y$ is the yield stress, $\dsK$ is a constant determined by the matrix composition, $n$ is a flow behavior index, $\Lambda$ is the nozzle diameter, and \rev{$U$} is the print speed.

The objective was to maximize the Oldroyd number and subsequently minimize the stress field around the nozzle. Increasing the volume fraction of FS in the gel resulted in stiffer gels, as evidenced by the increase in yield stress of the gels from 11 to 221 Pa (Figure \ref{fig:Printing}\textbf{c}), leading to improved shaping fidelity of the piezo ink. However, excessively high yield stress in the gel can cause static crevices that do not collapse or self-heal, resulting in print distortions. The self-healing and collapse of crevices depend on printing depth and the rheological properties of the support matrix. Gels with more than 4\% FS showed significant crevices and viscoplastic behavior. Therefore, a gel with 4\% FS, yielding a stress of 117 Pa, was chosen for its sufficient self-healing properties upon nozzle translation.

\subsubsection*{Printing speed design}
For the chosen piezoelectric ink of 43 vol\% and gel of 4 wt\% FS, the print speed was used as a final control parameter to minimize the yield region. Viscosity vs. shear rates for both the gel and ink were measured at typical printing speeds, showing a match in the shear rate region between 0.1 s\textsuperscript{-1} and 0.7 s\textsuperscript{-1} as seen in Figure \ref{fig:Printing}\textbf{d}.

The flow behavior index $n$ for these materials were then calculated by using the Oswald-de Waele power law for shear-thinning fluids 
\be
\eta=\dsK\dot{\gamma}^{n-1},
\ee
where, $\eta$ is the viscosity, $\dot{\gamma}$ [s\textsuperscript{-1}] is the shear rate. Using $n$ derived from the above equation, the range of printing speeds was determined based on the shear rate and nozzle diameter and was found to be between 2 mm/min and 50 mm/min. Low printing speeds (2 mm/min) minimized the yield region but resulted in higher print times and potential sedimentation and clogging of the piezo ink. \rev{Printing was conducted at two speeds: the bulk of the lattice (95\% of each strut's length) was printed at a higher speed (50 mm/min) within the matched viscosity regime, while the remaining 5\% near the nodes was printed at a lower speed (5 mm/min) to ensure minimal disturbance of the printed lines.} This strategy enabled relatively quick printing while ensuring well-connected and non-distorted lattices at the nodes, as shown in \figureautorefname~\ref{fig:Printing}\textbf{a}. Multiple lattices with different unit cells were then printed, as depicted in Figure \ref{fig:Printing}\textbf{e}. While our in-gel technique can print the truss structures with good accuracy, the complexity of the designs and sharp edges can influence the joints of struts. \rev{The oozing of ink from the nozzle during travel motions requires additional considerations. For the BCC and octet structures, it was sufficient to tune the printing speed and rheology to avoid unwanted \textit{spikes} in the direction of the travel, which originate from the pressure build-up in the nozzle.} \rev{This oozing of the ink can also be reduced by using a smaller particle size. However, this increases the viscosity of the ink and limits the overall volume fraction that can be achieved during printing. Other potential ways to avoid the overflow of the ink are implementing a sophisticated retraction mechanism for precise flow control or avoiding the use of gel altogether by simultaneous extrusion and curing of the ink during printing. We utilized a simple retraction mechanism by screw reversal for the optimized unit cell, where an excessive ink overflow was observed.} Next, we explore the piezoelectric response of the metamaterials, the directionality enabled by their design and compare their response with conventional bulk materials.

\subsection*{Tailored piezoelectricity of truss metamaterials}
\begin{figure}
\vspace{-16pt}
    \centering
    \includegraphics[width=\linewidth]{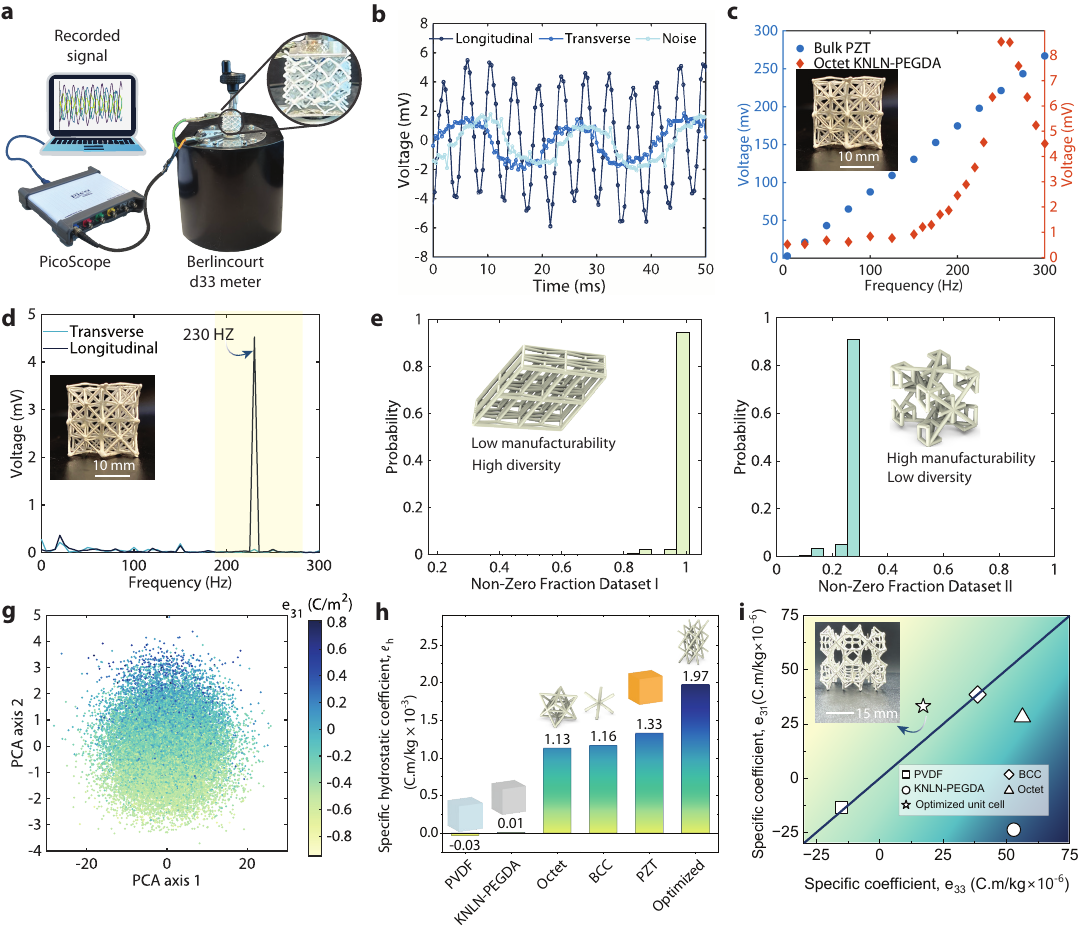}
    \captionsetup{justification=justified}
    \caption{\textbf{Evaluation of the tunable electromechanical response of piezoelectric truss metamaterials.} \textbf{a.} Schematic of the measurement setup used for recording the transient response of printed samples. The electrodes were fixed to the samples at the top and bottom surfaces. The force was applied either in the longitudinal direction or in the transverse direction, depending on the mode of operation. \textbf{b.} Transient voltage response of octet metamaterial in longitudinal and transverse modes under a dynamic load of 0.5 N. \textbf{c.} The measured frequency vs voltage response of an octet metamaterial and a bulk PZT sample. The bulk PZT sample shows a linear increase in voltage, while for octet samples, a peak is observed at 250 Hz. \textbf{d.} The frequency domain response of the octet samples under a harmonic force of 0.5 N at 230 Hz. The voltage peak at 230 Hz in longitudinal mode is found to be significantly higher than the transverse mode, as predicted by the computational homogenization. \textbf{e-f.} The comparison of the two design spaces utilized for optimization in terms of non-zero fraction, which is the ratio of number of non-zero components in the piezoelectric matrix to the total number of components. Design space I offers all 18 piezoelectric coefficients but lacks manufacturability due to extreme aspect ratios and skewed geometries. Design space II activates only five piezoelectric coefficients with a broader range and better manufacturability due to cubic symmetry. \textbf{g.} PCA performed in the latent space of dataset II, where with the increasing values of principal component 2, the value of $e_{31}$ increases. \textbf{h.} The comparison of specific hydrostatic piezoelectric coefficient of the optimized unit cell with bulk piezoelectric materials and BCC, and octet metamaterials. \textbf{i.} The $e_{31}$ vs $e_{32}$ plot for different piezoelectric materials shows that while most piezoelectric materials have $e_{33}$ higher than $e_{31}$ and lie below the 45\degree slope line, the ML-optimzied unit cell is able to reverse this trend and has higher $e_{31}$. \rev{The printed sample with the ML-optimized unit cell is shown in the inset.}}
    \label{fig:Measurements}
\end{figure}

In this section, we demonstrate the directionality of the piezoelectric response in truss metamaterials by testing 3D-printed $2\times$2$\times$2 tessellations of octet unit cells. To show the architecture-driven electromechanical responses of these samples, we measured them in longitudinal and transverse modes. A standard Berlincourt d\textsubscript{33}-meter was used to apply a harmonic force on the samples, and the generated voltage ($\phi$) was recorded using a pico logger, as shown in \figureautorefname~\ref{fig:Measurements}\textbf{a}. Multiple electromechanical phenomena, such as triboelectricity and flexoelectricity, can often co-exist in such systems and might influence accurate piezoelectric characterization. To eliminate the possible influence of triboelectricity due to contact-separation and sliding of different surfaces, all the measurements were conducted while maintaining a static force. Due to the macro scale size of the samples, the generated strain gradients are low, and thus the influence of flexoelectricity can be considered negligible \cite{sharma2025discontinuousgalerkinmethodbased}. With the robust measurement strategy, we turn to measure the directional response of the octet samples and show the directionality of response in piezoelectric truss metamaterials.

The amplitude of voltage in transverse and longitudinal modes is compared for the octet sample in \figureautorefname~\ref{fig:Measurements}\textbf{b}. While for the longitudinal mode, the force is applied to the electrodes, for measurements in the transverse mode, the samples are subjected to the harmonic excitation on the face normal to direction 1. The well-known 50 Hz noise is observed to have a significant magnitude compared to the piezoelectric signal. This noise can be easily eliminated since the measurements were performed at a forcing frequency of 230 Hz, as later shown in \figureautorefname~\ref{fig:Measurements}\textbf{d}. 

Due to lower resonant frequencies compared to bulk samples of similar dimensions, the piezoelectric response of piezoelectric truss metamaterials was observed to be highly sensitive to the forcing frequency. The voltage responses of an octet sample and a standard bulk PZT sample were measured at different frequencies in the range of 5-300 Hz, as shown in \figureautorefname~\ref{fig:Measurements}\textbf{c}. The PZT sample shows a linear increase in the voltage response with increasing forcing frequency, indicating that none of the resonance frequencies are close to the forcing frequencies. Thus, the equipment for measuring the piezoelectric coefficients of these samples are usually calibrated at these low frequencies for reliable estimates. For the octet sample, an exponential increase in voltage response with a peak at 250 Hz is observed, indicating mechanical resonance. This means that the harmonic force or impact measurements, commonly used for bulk piezoelectric samples, are not well-suited for characterizing the piezoelectric coefficients of the truss metamaterials. Therefore, to avoid the influence of mechanical resonance on characterization, we restrict ourselves to evaluating only the transient response of truss metamaterials in this low-frequency regime.

We further demonstrate the directionality of piezoelectric response in metamaterials by measuring the voltage response of octet samples in transverse and longitudinal modes. The frequency domain voltage response of the octet sample under a dynamic load of 230 Hz is compared in transverse and longitudinal modes in \figureautorefname~\ref{fig:Measurements}\textbf{d}. The voltage peak at 230 Hz in longitudinal mode is found to be approximately 65 times the peak in transverse mode due to the high $d_{33}/d_{31}$ ratio of the octet unit cells, in contrast to the base composite material, where this ratio is approximately 2. This shows the effectiveness of the truss metamaterials in providing tailored piezoelectric response. An important factor to be considered for the physical realization of this concept is the compromise between the tunable piezoelectric properties and manufacturability constraints.

The two design spaces presented in this study provide different combinations of manufacturability and range of properties. Design space I can offer more diverse piezoelectric responses compared to design space II. \figureautorefname~\ref{fig:Measurements}\textbf{e} shows the distribution of the fraction of non-zero coefficients of the piezoelectric matrix in design space I, which is defined as the ratio of number of non-zero piezoelectric coefficients in the piezoelectric matrix to the total number of coefficients. Most of the designs are found to have all 18 coefficients non-zero, indicated by a close to one probability distribution of the non-zero fraction of one. For design space II, most of the unit cells in the dataset are found to have a non-zero components fraction of 5/18, since only five of the 18 coefficients are non-zero, as shown in \figureautorefname~\ref{fig:Measurements}\textbf{f}. While limited to only five coefficients, all the unit cells in design space II have a cubic symmetry, leading to improved manufacturability. As shown in \figureautorefname~Supporting Information Figure S5\textbf{a}, the dataset from design space I shows a narrower range of coefficients $e_{31}$ and $e_{32}$, restricted to only positive values. The distribution of $e_{31}$ and $e_{32}$ for dataset II in \figureautorefname~Supporting Information Figure S5\textbf{b} shows these coefficients ranging from negative to positive values. Thus, the trade-off between manufacturability and range of properties should be considered depending upon the manufacturing capabilities available and the intended application. These distinct features of the two design spaces stem from the different parameterizations used to construct them.

We explored the two design spaces for the influence of specific design parameters on the unit cell designs and their piezoelectric properties. \figureautorefname~\ref{fig:Measurements}\textbf{g} shows the principal component analyses (PCA) conducted in the latent space of the VAE framework for design space II. Since the latent space was trained in conjunction with the property predictor NN, the latent space parameters are intertwined with the piezoelectric properties. A strong correlation between the principal component 2 and the piezoelectric coefficient $e_{31}$ was observed. The principal component 1 has a weaker influence on $e_{31}$, while the geometry still evolves along principal axis 1, as shown in Supporting Information Figure 7\textbf{a-b}, along with the evolution of the unit cell designs. Supporting Information Figure S7\textbf{c} shows PCA for the parameter space of design space I. We found that the principal components of design space I do not have a direct correlation with the piezoelectric coefficient $e_{31}$. This is due to the discrete nature of the design space I and small changes in design parameters resulting in sudden jumps in piezoelectric properties, as was also observed during the optimization in \figureautorefname~\ref{fig:ML}\textbf{c}. We take one example of optimization performed in each design space and compare their piezoelectricity with bulk materials.

We compared the piezoelectric performance of the metamaterials for application-relevant metrics. The optimized lattice obtained by maximizing the hydrostatic coefficient in design space I (Eq. \ref{eq:ds1_optim}) outperforms the commonly used bulk materials, as well as the KNLN-PEGDA composite. The optimized truss metamaterial offers a lightweight alternative for energy harvesting, actuation and sensing applications under hydrostatic loads and uniform actuation, such as in underwater sensing applications. In applications where unidirectional sensing or actuation is required, the isolation of modes of piezoelectricity becomes crucial. We explored one such possibility, where higher transverse piezoelectricity is desired compared to longitudinal piezoelectricity, by solving the optimization problem in Eq. \ref{eq:ds2_optim}. \figureautorefname~\ref{fig:Measurements}\textbf{i} shows the comparison of the optimized lattice with the bulk materials in the normalized $e_{33}$ vs $e_{31}$ landscape. \rev{The printed optimized lattice with $2\times2\times2$ tessellations is shown in the inset. While printing the optimized unit cell, the issue of oozing of the ink during travel motions, as discussed earlier, was found to be more pronounced. This is due to the smaller strut lengths and more travel motions compared to the simple structures shown in \figureautorefname~\ref{fig:Printing}\textbf{e}. We mitigated this by implementing a retraction step at the start of every travel motion, followed by an equal extrusion at the end of the travel move. for printing the optimized unit cell.} Most of the common bulk materials, as well as the base material, lie below the unit slope line since the magnitude of $e_{33}$ is higher than $e_{31}$ for these materials, PVDF being an exception with a slightly higher value of $e_{31}$. The optimized lattice shows the highest value of normalized $e_{31}-|e_{33}|$ and lies beyond the unit slope line. This enables the pure transverse mode operation of piezoelectric materials even under mixed loading conditions, with potential applications in specialized motion in micro-robots, medical equipment, and unidirectional electromechanical transducers.

\section*{Discussion}
Considering that most of the common bulk piezoelectric materials are restricted to the same symmetry class, our ML-based approach to designing new metamaterials offers clear advantages. Specifically, we achieved a hydrostatic piezoelectric coefficient for metamaterials higher than their bulk counterparts, and a significantly higher $e_{31}$ value compared to $e_{33}$. The optimized properties of these metamaterials extend beyond the training design space, demonstrating the extrapolatory capabilities of our proposed ML framework.

Through optimization, we also discovered that designs constrained within cubic symmetry are limited to the standard five piezoelectric coefficients. Accessing additional coefficients requires skewed geometries or aspect ratios in the unit cells. Consequently, our strategy of exploring two design spaces maximizes both absolute performance and anisotropy or manufacturability. Our results show that truss metamaterials offer exceptional flexibility in tuning the piezoelectric response, activating all 18 coefficients, and enabling rare or unique combinations of coefficients not found in conventional materials. Principal Component Analysis of the latent space of design space II and design space I revealed differing behaviors. In design space II, the second principal component directly influenced the magnitude of $e_{31}$, while design space I exhibited no discernible trend.

In characterizing the response of 3D-printed octet metamaterials in longitudinal mode, we observed an exponential increase in the piezoelectric response at low frequencies, peaking in local resonance at 250 Hz. This finding suggests that standard characterization frequencies and methods for bulk piezoelectric materials may not be suitable for lattice structures. We propose that future studies further explore the frequency dependence of piezoelectric materials. Despite these considerations, lightweight lattice structures optimized using our ML framework outperform bulk materials in key metrics, such as specific hydrostatic performance or achieving dominant transverse piezoelectricity over longitudinal modes. \rev{To demonstrate the concept of piezoelectric truss metamaterials, we 3D-printed and tested the octet topology. While our 3D printing technique provides a way to print the truss metamaterials, further advancements, such as improved flow control of the ink, are warranted to translate this concept to real-world applications.}
Nonetheless, our developed optimization framework and 3D-printing technique provide a novel way to design piezoelectric materials, overcoming limitations in material and symmetry classes. These advancements enable the creation of multi-modal electromechanical devices with tailored behaviors, potentially useful in applications such as micro robotics and health monitoring devices. Our electromechanical testing also reveals the importance of considering structural dynamics in individual structures and designing specific mechanical and electrical boundary conditions for lattice-based piezoelectric materials.

This work advances the field of piezoelectricity by demonstrating the potential to optimize piezoelectric performance through structural design, transcending the constraints of conventional bulk materials’ symmetries. Our proposed 3D-printing technique, coupled with a graph-based path planning framework, offers enhanced shaping freedom and precision, paving the way for advanced piezoelectric composite materials. These developments open new possibilities for electromechanical systems with engineered physical and responsive behaviors.

\section*{Methods}
\subsubsection*{Numerical homogenization}
For the base material used in simulations, isotropic elastic properties are assumed with Young's modulus $E^{\text{base}}= 60.606$ GPa and Poisson's ratio $\nu = 0.3$. For the piezoelectric properties, $e_{31}^{\text{base}}=e_{32}^{\text{base}}=-6.62281$ C/m$\textsuperscript{2}$, $e_{33}^{\text{base}}=23.2403$ C/m$\textsuperscript{2}$, and $e_{15}^{\text{base}}=e_{24}^{\text{base}}=17.0345$ C/m$\textsuperscript{2}$ are considered, while the relative dielectric permittivity is $\kappa_{11}=\kappa_{22}=\kappa_{33}=1433.6$.
\subsubsection*{Ceramic synthesis}
We synthesized the
K\textsubscript{0.485}Na\textsubscript{0.485}Li\textsubscript{0.03}NbO\textsubscript{3} (KNLN) ceramic powder through a conventional solid-state reaction with a two-step calcination process, reported previously\cite{james2016high}. Stoichiometric proportions of >99 \% Na\textsubscript{2}CO\textsubscript{3}, K\textsubscript{2}CO\textsubscript{3}, Li\textsubscript{2}CO\textsubscript{3}, and Nb\textsubscript{2}O\textsubscript{5} (Sigma Aldrich) were milled for at 200 rpm for 3 hours with 5 mm yttria-stabilized ZrO\textsubscript{2} balls in hexane. Afterward, the milled powders were air-dried on a hot plate at 100 \degree C for 1 hour, followed by calcination at 1050 \degree C for 3 hours with a heating rate of 5 \degree C min\textsuperscript{-1}. After the initial calcination, the powder was milled in IPA for 3 hours at 200 rpm, air-dried, and subjected to a second calcination at 925 \degree C for 20 hours, with a slower heating rate of 1 \degree C min\textsuperscript{-1}. The resulting calcined powder was ball-milled again in IPA for 15 minutes, air-dried, sieved through a 90 µm mesh, and stored under vacuum. Particle morphology was examined using a Scanning Electron Microscope (SEM; JEOL JSM-7500F, Nieuw Vennep, Netherlands), and the particle size distribution was determined using a Malvern Mastersizer 3000. The analysis was performed on a 0.1\% weight/volume aqueous suspension of the powders with sodium dodecyl sulfate as a surfactant. The mean value of the distribution represents the particle size of each type of powder.
\subsubsection*{Ink Preparation} The synthesized ceramic powder was mixed with the UV-sensitive monomer poly(ethylene glycol) diacrylate with Mn 700 (Sigma Aldrich) in varying volume fractions. This mixture was combined using a planetary mixer (SpeedMixer DAC 150.1 FVZ) at 3500 rpm for 2 minutes, with 30-second intervals and a 15-second break to prevent excessive heating. Two photoinitiators, Irgacure 819 and 184, were added to maximize curing depth due to the high ceramic filler content, creating a UV-curable piezoelectric ink. The photoinitiators were added to the KNLN-PEGDA mixture in ratios of 1\% and 2\% by weight relative to the monomer and mixed at 1000 rpm for 5 minutes to ensure homogeneous distribution.
\subsubsection*{Support gel preparation} The gel is created by dispersing 4 wt\% of fumed silica (Sigma Aldrich) in silicone oil. This mixture is processed in a planetary mixer (SpeedMixer DAC 150.1 FVZ) at 3500 rpm for 5 minutes to ensure thorough dispersion and uniformity.
\subsubsection*{Rheological characterization}
The rheological properties of the inks were characterized using a rotational rheometer (Haake Mars III, Thermoscientific) with a 20 mm diameter serrated plate geometry using a gap height of 1 mm. Serrated plates prevent artifacts in measurements arising from wall slip, which has been studied to occur when highly loaded dispersions are subject to high localized deformation. Shear storage and loss moduli were determined as a function of shear strain via dynamic amplitude sweeps at a fixed frequency of 1 Hz with a stress sweep. The yield stress was measured via steady-state flow experiments with a sweep of shear stress and measuring. We used the point of change in slope of the log-log plot of shear stress vs strain to calculate the yield stress of the ink. The viscosity vs shear rates was measured between shear rates from 0.1 to 100 s$^{-1}$.
\subsubsection*{Printing and curing}
A commercial desktop printer (Ultimaker 2+) was modified by replacing the original print head with a custom-made ink extrusion system consisting of a 25ml syringe holder and a mechanically driven syringe pump. The inks loaded into syringes were attached with stainless steel nozzles of 1.5-inch length with an inner diameter of 0.5 mm and were used for printing. The printed lines or parts were cured using an Omnicure S1500 (Lumen Dynamics) UV-lamp at 100\% intensity at 20 cm for 30 seconds for the final part.
\subsubsection*{Poling}
The printed samples are first cleaned in a Hexane solution to wash away the silicone oil and facilitate better electrode adhesion. A corona poling setup is used to pole the sample with 5 kV as the bottom plate potential and 50 kV applied to the pin electrodes. The base plate is heated to 120 \degree C to allow for the reorientation of the dipoles in the samples. Samples are poled for 6 hours, after which they are cooled to room temperature with the electric field applied and used for measurements.
\subsubsection*{Piezoelectric characterization}
Aluminum metal electrodes were attached to the faces, which were in the same direction as the poling direction (3), using a silver epoxy adhesive. The piezoelectric voltage was measured using a Berlincourt (PM 300, Piezo test, London, UK) piezometer with a static force of 2 N and a dynamic peak-to-peak sinusoidal excitation ranging between 0.05 N and 0.5 N at 110 Hz. The output voltage was measured using a Picoscope (5442D series).

\section*{Data availability}
The dataset used for training the ML frameworks in this work is available at the \href{https://doi.org/10.5281/zenodo.17041483}{repository}\cite{sharma_2025_17041483}. 

\section*{Code availability}
The codes used for homogenization, ML training and optimization, and path planning are available on \href{https://doi.org/10.5281/zenodo.17041546}{repository}\cite{sharma_2025_17041546}, and \href{https://github.com/mmc-group/piezoelectric-truss-metamaterials-design}{https://github.com/mmc-group/piezoelectric-truss-metamaterials-design}.

\section*{Acknowledgments}
S.S. and K.M. acknowledge financial support from the European Union (ERC CoG, AM-IMATE, 101088968). S.S. acknowledges financial support from the European Commission (HORIZON-MSCA-2023-PF-01, Meta-SMART, 101153507). We also thank Li Zheng for insightful discussions during the development of the VAE framework. 

\section*{Author contributions}
\textbf{S.S.} and \textbf{S.K.} developed the piezoelectric truss homogenization code. \textbf{S.S.} and \textbf{P.T}. trained the machine learning models and performed the design optimization. \textbf{S.S.} developed the path planning code for printing, \textbf{S.K.A.} performed the direct ink writing, and developed gel and ink formulation. \textbf{S.S.} and \textbf{S.K.A.} built the experimental setup for characterizing the fabricated structures and performed the measurements. \textbf{ S.S.}, \textbf{J.J.}, \textbf{K.M.}, and \textbf{S.K.} conceptualized the research. \textbf{J.J.}, \textbf{K.M.}, and \textbf{S.K.} supervised the research. \textbf{S.S.}, \textbf{S.K.A.}, and \textbf{P.T.} wrote the initial draft, and all authors contributed to editing and revising the manuscript.
\section*{Competing interests}
The authors declare no competing interests.

\bibliography{Bib}

\begin{thebibliography}{10}
\urlstyle{rm}
\expandafter\ifx\csname url\endcsname\relax
  \def\url#1{\texttt{#1}}\fi
\expandafter\ifx\csname urlprefix\endcsname\relax\def\urlprefix{URL }\fi
\expandafter\ifx\csname doiprefix\endcsname\relax\def\doiprefix{DOI: }\fi
\providecommand{\bibinfo}[2]{#2}
\providecommand{\eprint}[2][]{\url{#2}}

\bibitem{kim2020biomolecular}
\bibinfo{author}{Kim, D.} \emph{et~al.}
\newblock \bibinfo{journal}{\bibinfo{title}{Biomolecular piezoelectric
  materials: from amino acids to living tissues}}.
\newblock {\emph{\JournalTitle{Advanced Materials}}}
  \textbf{\bibinfo{volume}{32}}, \bibinfo{pages}{1906989}
  (\bibinfo{year}{2020}).

\bibitem{tandon2018piezoelectric}
\bibinfo{author}{Tandon, B.}, \bibinfo{author}{Blaker, J.~J.} \&
  \bibinfo{author}{Cartmell, S.~H.}
\newblock \bibinfo{journal}{\bibinfo{title}{Piezoelectric materials as
  stimulatory biomedical materials and scaffolds for bone repair}}.
\newblock {\emph{\JournalTitle{Acta biomaterialia}}}
  \textbf{\bibinfo{volume}{73}}, \bibinfo{pages}{1--20} (\bibinfo{year}{2018}).

\bibitem{xu2021construction}
\bibinfo{author}{Xu, Q.} \emph{et~al.}
\newblock \bibinfo{journal}{\bibinfo{title}{Construction of bio-piezoelectric
  platforms: From structures and synthesis to applications}}.
\newblock {\emph{\JournalTitle{Advanced Materials}}}
  \textbf{\bibinfo{volume}{33}}, \bibinfo{pages}{2008452}
  (\bibinfo{year}{2021}).

\bibitem{boukabache2014toward}
\bibinfo{author}{Boukabache, H.}, \bibinfo{author}{Escriba, C.} \&
  \bibinfo{author}{Fourniols, J.-Y.}
\newblock \bibinfo{journal}{\bibinfo{title}{Toward smart aerospace structures:
  Design of a piezoelectric sensor and its analog interface for flaw
  detection}}.
\newblock {\emph{\JournalTitle{Sensors}}} \textbf{\bibinfo{volume}{14}},
  \bibinfo{pages}{20543--20561} (\bibinfo{year}{2014}).

\bibitem{elahi2018response}
\bibinfo{author}{Elahi, H.} \emph{et~al.}
\newblock \bibinfo{journal}{\bibinfo{title}{Response of piezoelectric materials
  on thermomechanical shocking and electrical shocking for aerospace
  applications}}.
\newblock {\emph{\JournalTitle{Microsystem Technologies}}}
  \textbf{\bibinfo{volume}{24}}, \bibinfo{pages}{3791--3798}
  (\bibinfo{year}{2018}).

\bibitem{jackson2013flexible}
\bibinfo{author}{Jackson, N.}, \bibinfo{author}{Keeney, L.} \&
  \bibinfo{author}{Mathewson, A.}
\newblock \bibinfo{journal}{\bibinfo{title}{Flexible-cmos and biocompatible
  piezoelectric aln material for mems applications}}.
\newblock {\emph{\JournalTitle{Smart Materials and Structures}}}
  \textbf{\bibinfo{volume}{22}}, \bibinfo{pages}{115033}
  (\bibinfo{year}{2013}).

\bibitem{kim2012piezoelectric}
\bibinfo{author}{Kim, S.-G.}, \bibinfo{author}{Priya, S.} \&
  \bibinfo{author}{Kanno, I.}
\newblock \bibinfo{journal}{\bibinfo{title}{Piezoelectric mems for energy
  harvesting}}.
\newblock {\emph{\JournalTitle{MRS bulletin}}} \textbf{\bibinfo{volume}{37}},
  \bibinfo{pages}{1039--1050} (\bibinfo{year}{2012}).

\bibitem{kim2018optimized}
\bibinfo{author}{Kim, K.-B.} \emph{et~al.}
\newblock \bibinfo{journal}{\bibinfo{title}{Optimized composite piezoelectric
  energy harvesting floor tile for smart home energy management}}.
\newblock {\emph{\JournalTitle{Energy Conversion and Management}}}
  \textbf{\bibinfo{volume}{171}}, \bibinfo{pages}{31--37}
  (\bibinfo{year}{2018}).

\bibitem{yang2018preliminary}
\bibinfo{author}{Yang, H.}, \bibinfo{author}{Wang, L.}, \bibinfo{author}{Zhou,
  B.}, \bibinfo{author}{Wei, Y.} \& \bibinfo{author}{Zhao, Q.}
\newblock \bibinfo{journal}{\bibinfo{title}{A preliminary study on the highway
  piezoelectric power supply system}}.
\newblock {\emph{\JournalTitle{International Journal of Pavement Research and
  Technology}}} \textbf{\bibinfo{volume}{11}}, \bibinfo{pages}{168--175}
  (\bibinfo{year}{2018}).

\bibitem{qian2020piezoelectric}
\bibinfo{author}{Qian, W.}, \bibinfo{author}{Yang, W.}, \bibinfo{author}{Zhang,
  Y.}, \bibinfo{author}{Bowen, C.~R.} \& \bibinfo{author}{Yang, Y.}
\newblock \bibinfo{journal}{\bibinfo{title}{Piezoelectric materials for
  controlling electro-chemical processes}}.
\newblock {\emph{\JournalTitle{Nano-Micro Letters}}}
  \textbf{\bibinfo{volume}{12}}, \bibinfo{pages}{1--39} (\bibinfo{year}{2020}).

\bibitem{hu2019improved}
\bibinfo{author}{Hu, X.} \emph{et~al.}
\newblock \bibinfo{journal}{\bibinfo{title}{Improved piezoelectric sensing
  performance of p (vdf--trfe) nanofibers by utilizing bto nanoparticles and
  penetrated electrodes}}.
\newblock {\emph{\JournalTitle{ACS applied materials \& interfaces}}}
  \textbf{\bibinfo{volume}{11}}, \bibinfo{pages}{7379--7386}
  (\bibinfo{year}{2019}).

\bibitem{shi2021interface}
\bibinfo{author}{Shi, K.} \emph{et~al.}
\newblock \bibinfo{journal}{\bibinfo{title}{Interface induced performance
  enhancement in flexible batio3/pvdf-trfe based piezoelectric
  nanogenerators}}.
\newblock {\emph{\JournalTitle{Nano Energy}}} \textbf{\bibinfo{volume}{80}},
  \bibinfo{pages}{105515} (\bibinfo{year}{2021}).

\bibitem{abbasipour2019improving}
\bibinfo{author}{Abbasipour, M.} \emph{et~al.}
\newblock \bibinfo{journal}{\bibinfo{title}{Improving piezoelectric and
  pyroelectric properties of electrospun pvdf nanofibers using nanofillers for
  energy harvesting application}}.
\newblock {\emph{\JournalTitle{Polymers for Advanced Technologies}}}
  \textbf{\bibinfo{volume}{30}}, \bibinfo{pages}{279--291}
  (\bibinfo{year}{2019}).

\bibitem{krishnaswamy2019improving}
\bibinfo{author}{Krishnaswamy, J.~A.} \emph{et~al.}
\newblock \bibinfo{journal}{\bibinfo{title}{Improving the performance of
  lead-free piezoelectric composites by using polycrystalline inclusions and
  tuning the dielectric matrix environment}}.
\newblock {\emph{\JournalTitle{Smart Materials and Structures}}}
  \textbf{\bibinfo{volume}{28}}, \bibinfo{pages}{075032}
  (\bibinfo{year}{2019}).

\bibitem{chen2018y2o3}
\bibinfo{author}{Chen, Z.-h.} \emph{et~al.}
\newblock \bibinfo{journal}{\bibinfo{title}{Y2o3 doped ba0. 9ca0. 1ti0. 9sn0.
  1o3 ceramics with improved piezoelectric properties}}.
\newblock {\emph{\JournalTitle{Journal of the European Ceramic Society}}}
  \textbf{\bibinfo{volume}{38}}, \bibinfo{pages}{1349--1355}
  (\bibinfo{year}{2018}).

\bibitem{wang2018optimization}
\bibinfo{author}{Wang, C.}, \bibinfo{author}{Zhao, J.}, \bibinfo{author}{Li,
  Q.} \& \bibinfo{author}{Li, Y.}
\newblock \bibinfo{journal}{\bibinfo{title}{Optimization design and
  experimental investigation of piezoelectric energy harvesting devices for
  pavement}}.
\newblock {\emph{\JournalTitle{Applied energy}}}
  \textbf{\bibinfo{volume}{229}}, \bibinfo{pages}{18--30}
  (\bibinfo{year}{2018}).

\bibitem{de2019topology}
\bibinfo{author}{De~Almeida, B.~V.}, \bibinfo{author}{Cunha, D.~C.} \&
  \bibinfo{author}{Pavanello, R.}
\newblock \bibinfo{journal}{\bibinfo{title}{Topology optimization of bimorph
  piezoelectric energy harvesters considering variable electrode location}}.
\newblock {\emph{\JournalTitle{Smart Materials and Structures}}}
  \textbf{\bibinfo{volume}{28}}, \bibinfo{pages}{085030}
  (\bibinfo{year}{2019}).

\bibitem{yoon2018multiphysics}
\bibinfo{author}{Yoon, G.~H.}, \bibinfo{author}{Choi, H.} \&
  \bibinfo{author}{Hur, S.}
\newblock \bibinfo{journal}{\bibinfo{title}{Multiphysics topology optimization
  for piezoelectric acoustic focuser}}.
\newblock {\emph{\JournalTitle{Computer Methods in Applied Mechanics and
  Engineering}}} \textbf{\bibinfo{volume}{332}}, \bibinfo{pages}{600--623}
  (\bibinfo{year}{2018}).

\bibitem{katsouras2016negative}
\bibinfo{author}{Katsouras, I.} \emph{et~al.}
\newblock \bibinfo{journal}{\bibinfo{title}{The negative piezoelectric effect
  of the ferroelectric polymer poly (vinylidene fluoride)}}.
\newblock {\emph{\JournalTitle{Nature materials}}}
  \textbf{\bibinfo{volume}{15}}, \bibinfo{pages}{78--84}
  (\bibinfo{year}{2016}).

\bibitem{xu20163d}
\bibinfo{author}{Xu, Q.} \emph{et~al.}
\newblock \bibinfo{journal}{\bibinfo{title}{3d-printing of inverted pyramid
  suspending architecture for pyroelectric infrared detectors with inhibited
  microphonic effect}}.
\newblock {\emph{\JournalTitle{Infrared Physics \& Technology}}}
  \textbf{\bibinfo{volume}{76}}, \bibinfo{pages}{111--115}
  (\bibinfo{year}{2016}).

\bibitem{chong2002pyroelectric}
\bibinfo{author}{Chong, N.}, \bibinfo{author}{Chan, H.} \&
  \bibinfo{author}{Choy, C.}
\newblock \bibinfo{journal}{\bibinfo{title}{Pyroelectric sensor array for
  in-line monitoring of infrared laser}}.
\newblock {\emph{\JournalTitle{Sensors and Actuators A: Physical}}}
  \textbf{\bibinfo{volume}{96}}, \bibinfo{pages}{231--238}
  (\bibinfo{year}{2002}).

\bibitem{bastek2022inverting}
\bibinfo{author}{Bastek, J.-H.}, \bibinfo{author}{Kumar, S.},
  \bibinfo{author}{Telgen, B.}, \bibinfo{author}{Glaesener, R.~N.} \&
  \bibinfo{author}{Kochmann, D.~M.}
\newblock \bibinfo{journal}{\bibinfo{title}{Inverting the structure--property
  map of truss metamaterials by deep learning}}.
\newblock {\emph{\JournalTitle{Proceedings of the National Academy of
  Sciences}}} \textbf{\bibinfo{volume}{119}}, \bibinfo{pages}{e2111505119}
  (\bibinfo{year}{2022}).

\bibitem{zheng2023unifying}
\bibinfo{author}{Zheng, L.}, \bibinfo{author}{Karapiperis, K.},
  \bibinfo{author}{Kumar, S.} \& \bibinfo{author}{Kochmann, D.~M.}
\newblock \bibinfo{journal}{\bibinfo{title}{Unifying the design space and
  optimizing linear and nonlinear truss metamaterials by generative modeling}}.
\newblock {\emph{\JournalTitle{Nature Communications}}}
  \textbf{\bibinfo{volume}{14}}, \bibinfo{pages}{7563} (\bibinfo{year}{2023}).

\bibitem{shaikeea2022toughness}
\bibinfo{author}{Shaikeea, A. J.~D.}, \bibinfo{author}{Cui, H.},
  \bibinfo{author}{O’Masta, M.}, \bibinfo{author}{Zheng, X.~R.} \&
  \bibinfo{author}{Deshpande, V.~S.}
\newblock \bibinfo{journal}{\bibinfo{title}{The toughness of mechanical
  metamaterials}}.
\newblock {\emph{\JournalTitle{Nature Materials}}}
  \textbf{\bibinfo{volume}{21}}, \bibinfo{pages}{297--304}
  (\bibinfo{year}{2022}).

\bibitem{bertoldi2017flexible}
\bibinfo{author}{Bertoldi, K.}, \bibinfo{author}{Vitelli, V.},
  \bibinfo{author}{Christensen, J.} \& \bibinfo{author}{Van~Hecke, M.}
\newblock \bibinfo{journal}{\bibinfo{title}{Flexible mechanical
  metamaterials}}.
\newblock {\emph{\JournalTitle{Nature Reviews Materials}}}
  \textbf{\bibinfo{volume}{2}}, \bibinfo{pages}{1--11} (\bibinfo{year}{2017}).

\bibitem{zheng2014ultralight}
\bibinfo{author}{Zheng, X.} \emph{et~al.}
\newblock \bibinfo{journal}{\bibinfo{title}{Ultralight, ultrastiff mechanical
  metamaterials}}.
\newblock {\emph{\JournalTitle{Science}}} \textbf{\bibinfo{volume}{344}},
  \bibinfo{pages}{1373--1377} (\bibinfo{year}{2014}).

\bibitem{rafsanjani2015snapping}
\bibinfo{author}{Rafsanjani, A.}, \bibinfo{author}{Akbarzadeh, A.} \&
  \bibinfo{author}{Pasini, D.}
\newblock \bibinfo{journal}{\bibinfo{title}{Snapping mechanical metamaterials
  under tension}}.
\newblock {\emph{\JournalTitle{Advanced Materials}}}
  \textbf{\bibinfo{volume}{27}}, \bibinfo{pages}{5931--5935}
  (\bibinfo{year}{2015}).

\bibitem{rocklin2017transformable}
\bibinfo{author}{Rocklin, D.}, \bibinfo{author}{Zhou, S.},
  \bibinfo{author}{Sun, K.} \& \bibinfo{author}{Mao, X.}
\newblock \bibinfo{journal}{\bibinfo{title}{Transformable topological
  mechanical metamaterials}}.
\newblock {\emph{\JournalTitle{Nature communications}}}
  \textbf{\bibinfo{volume}{8}}, \bibinfo{pages}{1--9} (\bibinfo{year}{2017}).

\bibitem{dudek2022micro}
\bibinfo{author}{Dudek, K.~K.}, \bibinfo{author}{Mart{\'\i}nez, J. A.~I.},
  \bibinfo{author}{Ulliac, G.} \& \bibinfo{author}{Kadic, M.}
\newblock \bibinfo{journal}{\bibinfo{title}{Micro-scale auxetic hierarchical
  mechanical metamaterials for shape morphing}}.
\newblock {\emph{\JournalTitle{Advanced Materials}}}
  \textbf{\bibinfo{volume}{34}}, \bibinfo{pages}{2110115}
  (\bibinfo{year}{2022}).

\bibitem{cui2019three}
\bibinfo{author}{Cui, H.} \emph{et~al.}
\newblock \bibinfo{journal}{\bibinfo{title}{Three-dimensional printing of
  piezoelectric materials with designed anisotropy and directional response}}.
\newblock {\emph{\JournalTitle{Nature materials}}}
  \textbf{\bibinfo{volume}{18}}, \bibinfo{pages}{234--241}
  (\bibinfo{year}{2019}).

\bibitem{cui2022design}
\bibinfo{author}{Cui, H.} \emph{et~al.}
\newblock \bibinfo{journal}{\bibinfo{title}{Design and printing of
  proprioceptive three-dimensional architected robotic metamaterials}}.
\newblock {\emph{\JournalTitle{Science}}} \textbf{\bibinfo{volume}{376}},
  \bibinfo{pages}{1287--1293} (\bibinfo{year}{2022}).

\bibitem{shi20243d}
\bibinfo{author}{Shi, J.} \emph{et~al.}
\newblock \bibinfo{journal}{\bibinfo{title}{3d printed architected shell-based
  ferroelectric metamaterials with programmable piezoelectric and pyroelectric
  properties}}.
\newblock {\emph{\JournalTitle{Nano Energy}}} \textbf{\bibinfo{volume}{123}},
  \bibinfo{pages}{109385} (\bibinfo{year}{2024}).

\bibitem{ammu20243d}
\bibinfo{author}{Ammu, S.~K.} \emph{et~al.}
\newblock \bibinfo{journal}{\bibinfo{title}{3d printing of lead-free
  piezoelectric ultrasound transducers}}.
\newblock {\emph{\JournalTitle{Advanced Materials Technologies}}}
  \bibinfo{pages}{2400858} (\bibinfo{year}{2024}).

\bibitem{tao2022multi}
\bibinfo{author}{Tao, R.}, \bibinfo{author}{Granier, F.} \&
  \bibinfo{author}{Therriault, D.}
\newblock \bibinfo{journal}{\bibinfo{title}{Multi-material freeform 3d printing
  of flexible piezoelectric composite sensors using a supporting fluid}}.
\newblock {\emph{\JournalTitle{Additive Manufacturing}}}
  \textbf{\bibinfo{volume}{60}}, \bibinfo{pages}{103243}
  (\bibinfo{year}{2022}).

\bibitem{sharma_design_2021}
\bibinfo{author}{Sharma, S.}, \bibinfo{author}{Kumar, R.},
  \bibinfo{author}{Talha, M.} \& \bibinfo{author}{Vaish, R.}
\newblock \bibinfo{journal}{\bibinfo{title}{Design of spatially varying
  electrical poling for enhanced piezoelectricity in
  {Pb}({Mg1}/{3Nb2}/3){O3}–0.{35PbTiO3}}}.
\newblock {\emph{\JournalTitle{International Journal of Mechanics and Materials
  in Design}}} \textbf{\bibinfo{volume}{17}}, \bibinfo{pages}{99--118},
  \doiprefix\url{10.1007/s10999-020-09514-w} (\bibinfo{year}{2021}).

\bibitem{zok2016periodic}
\bibinfo{author}{Zok, F.~W.}, \bibinfo{author}{Latture, R.~M.} \&
  \bibinfo{author}{Begley, M.~R.}
\newblock \bibinfo{journal}{\bibinfo{title}{Periodic truss structures}}.
\newblock {\emph{\JournalTitle{Journal of the Mechanics and Physics of
  Solids}}} \textbf{\bibinfo{volume}{96}}, \bibinfo{pages}{184--203}
  (\bibinfo{year}{2016}).

\bibitem{weeks2023embedded}
\bibinfo{author}{Weeks, R.~D.}, \bibinfo{author}{Truby, R.~L.},
  \bibinfo{author}{Uzel, S.~G.} \& \bibinfo{author}{Lewis, J.~A.}
\newblock \bibinfo{journal}{\bibinfo{title}{Embedded 3d printing of
  multimaterial polymer lattices via graph-based print path planning}}.
\newblock {\emph{\JournalTitle{Advanced Materials}}}
  \textbf{\bibinfo{volume}{35}}, \bibinfo{pages}{2206958}
  (\bibinfo{year}{2023}).

\bibitem{pairam2013stable}
\bibinfo{author}{Pairam, E.} \emph{et~al.}
\newblock \bibinfo{journal}{\bibinfo{title}{Stable nematic droplets with
  handles}}.
\newblock {\emph{\JournalTitle{Proceedings of the National Academy of
  Sciences}}} \textbf{\bibinfo{volume}{110}}, \bibinfo{pages}{9295--9300}
  (\bibinfo{year}{2013}).

\bibitem{grosskopf2018viscoplastic}
\bibinfo{author}{Grosskopf, A.~K.} \emph{et~al.}
\newblock \bibinfo{journal}{\bibinfo{title}{Viscoplastic matrix materials for
  embedded 3d printing}}.
\newblock {\emph{\JournalTitle{ACS applied materials \& interfaces}}}
  \textbf{\bibinfo{volume}{10}}, \bibinfo{pages}{23353--23361}
  (\bibinfo{year}{2018}).

\bibitem{sharma2025discontinuousgalerkinmethodbased}
\bibinfo{author}{Sharma, S.}, \bibinfo{author}{Anitescu, C.} \&
  \bibinfo{author}{Rabczuk, T.}
\newblock \bibinfo{journal}{\bibinfo{title}{A discontinuous galerkin method
  based isogeometric analysis framework for flexoelectricity in
  micro-architected dielectric solids}}.
\newblock {\emph{\JournalTitle{Computers \& Structures}}}
  \textbf{\bibinfo{volume}{308}}, \bibinfo{pages}{107641},
  \doiprefix\url{https://doi.org/10.1016/j.compstruc.2024.107641}
  (\bibinfo{year}{2025}).

\bibitem{james2016high}
\bibinfo{author}{James, N.~K.}, \bibinfo{author}{Deutz, D.~B.},
  \bibinfo{author}{Bose, R.~K.}, \bibinfo{author}{van~der Zwaag, S.} \&
  \bibinfo{author}{Groen, P.}
\newblock \bibinfo{journal}{\bibinfo{title}{High piezoelectric voltage
  coefficient in structured lead-free (k, na, li) nbo3 particulate—epoxy
  composites}}.
\newblock {\emph{\JournalTitle{Journal of the American Ceramic Society}}}
  \textbf{\bibinfo{volume}{99}}, \bibinfo{pages}{3957--3963}
  (\bibinfo{year}{2016}).

\bibitem{sharma_2025_17041483}
\bibinfo{author}{Sharma, S.}
\newblock \bibinfo{title}{Dataset underlying the publication "piezoelectric
  truss metamaterials: data-driven design and additive manufacturing"},
  \doiprefix\url{10.5281/zenodo.17041483} (\bibinfo{year}{2025}).

\bibitem{sharma_2025_17041546}
\bibinfo{author}{Sharma, S.}
\newblock \bibinfo{title}{piezoelectric-truss-metamaterials-design},
  \doiprefix\url{10.5281/zenodo.17041546} (\bibinfo{year}{2025}).

\end{thebibliography}


\begin{thebibliography}{1}
\urlstyle{rm}
\expandafter\ifx\csname url\endcsname\relax
  \def\url#1{\texttt{#1}}\fi
\expandafter\ifx\csname urlprefix\endcsname\relax\def\urlprefix{URL }\fi
\expandafter\ifx\csname doiprefix\endcsname\relax\def\doiprefix{DOI: }\fi
\providecommand{\bibinfo}[2]{#2}
\providecommand{\eprint}[2][]{\url{#2}}

\bibitem{zhang2017novel}
\bibinfo{author}{Zhang, Y.}, \bibinfo{author}{Shang, S.} \&
  \bibinfo{author}{Liu, S.}
\newblock \bibinfo{journal}{\bibinfo{title}{A novel implementation algorithm of
  asymptotic homogenization for predicting the effective coefficient of thermal
  expansion of periodic composite materials}}.
\newblock {\emph{\JournalTitle{Acta Mechanica Sinica}}}
  \textbf{\bibinfo{volume}{33}}, \bibinfo{pages}{368--381}
  (\bibinfo{year}{2017}).

\bibitem{yang2004numerical}
\bibinfo{author}{Yang, Q.-S.} \& \bibinfo{author}{Becker, W.}
\newblock \bibinfo{journal}{\bibinfo{title}{Numerical investigation for stress,
  strain and energy homogenization of orthotropic composite with periodic
  microstructure and non-symmetric inclusions}}.
\newblock {\emph{\JournalTitle{Computational materials science}}}
  \textbf{\bibinfo{volume}{31}}, \bibinfo{pages}{169--180}
  (\bibinfo{year}{2004}).

\bibitem{lumpe2021exploring}
\bibinfo{author}{Lumpe, T.~S.} \& \bibinfo{author}{Stankovic, T.}
\newblock \bibinfo{journal}{\bibinfo{title}{Exploring the property space of
  periodic cellular structures based on crystal networks}}.
\newblock {\emph{\JournalTitle{Proceedings of the National Academy of
  Sciences}}} \textbf{\bibinfo{volume}{118}}, \bibinfo{pages}{e2003504118}
  (\bibinfo{year}{2021}).

\bibitem{kingma2014adam}
\bibinfo{author}{Kingma, D.~P.}
\newblock \bibinfo{journal}{\bibinfo{title}{Adam: A method for stochastic
  optimization}}.
\newblock {\emph{\JournalTitle{arXiv preprint arXiv:1412.6980}}}
  (\bibinfo{year}{2014}).

\bibitem{zheng2023unifying}
\bibinfo{author}{Zheng, L.}, \bibinfo{author}{Karapiperis, K.},
  \bibinfo{author}{Kumar, S.} \& \bibinfo{author}{Kochmann, D.~M.}
\newblock \bibinfo{journal}{\bibinfo{title}{Unifying the design space and
  optimizing linear and nonlinear truss metamaterials by generative modeling}}.
\newblock {\emph{\JournalTitle{Nature Communications}}}
  \textbf{\bibinfo{volume}{14}}, \bibinfo{pages}{7563} (\bibinfo{year}{2023}).

\end{thebibliography}

\end{document}

\flushbottom
\maketitle

\renewcommand\thefigure{S\arabic{figure}}   
\renewcommand\thetable{S\arabic{table}}   
\tableofcontents
\section{Numerical homogenization}
To compute the effective electromechanical properties of the piezoelectric metamaterials, we developed an in-house code based on Bernoulli-Euler beam elements. We employ two-noded elements with seven degrees of freedom (dofs) per node, including six mechanical dofs and electric potential as the electrical dof. We extend the asymptotic homogenization framework developed by Zhang et al. \cite{zhang2017novel} for elastic properties to compute the effective electromechanical properties of the piezoelectric metamaterials. After assembling the stiffness matrix of the unit cell, the global FE equation can be written as
\be
\label{eq:FEeq}
   \bfK \bfU= \bff,
\ee
where $\bfK$ is the global stiffness matrix, $\bfU$ is the displacement vector, and $\bff$ is the force vector corresponding to mechanical and electric dofs. The nodal dofs are ordered as $\{u_x \ u_y \ u_z \ \theta_x \ \theta_y  \ \theta_z \ \phi\}^T$, where $u$, and $\theta$ are the displacement and rotations, and $\phi$ is the nodal electric potential. Since the poling direction is considered to be along Z axis, the piezoelectric tensor $\bfe^{\text{base}}$ is first transformed to local coordinates (x, y, z) using the transformation matrix $\bfR$ written as
\be
\label{eq:Rmat}
\bfR = 
\begin{bmatrix}
    \beta_{Xx} & \beta_{Yx} & \beta_{Zx} \\
    \beta_{Xy} & \beta_{Yy} & \beta_{Zy} \\
    \beta_{Zx} & \beta_{Zy} & \beta_{Zz}
\end{bmatrix},
\ee
where $\beta_{Xx}$, $\beta_{Yx}$, and  $\beta_{Zx}$ are the direction cosines of the $x$-axis, with respect to global X, Y, and Z axes, respectively, and the similar notation is followed for the second and third rows of the matrix. The transformed $\bfe^{\text{base}}$ is then used to compute the electromechanical FE stiffness matrix in the struts' local coordinates $xyz$.

Since all the unit cells generated in both the datasets have periodic geometry, the corners, opposing faces, and parallel edges can be coupled using a transformation matrix $\bfT$ as $\bfU=\bfT\Tilde{\bfU}$. Where $\Tilde{\bfU}$ is the reduced displacement vector under periodic boundary conditions (PBCs). The expanded matrix form of this transformation can be written as \cite{yang2004numerical, lumpe2021exploring}

\be
\begin{bmatrix}
    U_{C1} \\ U_{C2} \\ U_{C3} \\ U_{C4} \\ U_{C5} \\ U_{C6} \\ U_{C7} \\ U_{C8} \\ U_{E1} \\ U_{E2} \\ U_{E3} \\ U_{E4} \\ U_{E5} \\ U_{E6} \\ U_{E7} \\ U_{E8} \\ U_{E9} \\ U_{E10} \\ U_{E11} \\ U_{E12} \\ U_{F1} \\ U_{F2} \\ U_{F3} \\ U_{F4} \\ U_{F5} \\ U_{F6} \\ U_{int}
\end{bmatrix}_{n\times1}
= \ \
\begin{bmatrix}
   I & 0 & 0 & 0 & 0 & 0 & 0 & 0 \\
   I & 0 & 0 & 0 & 0 & 0 & 0 & 0 \\
   I & 0 & 0 & 0 & 0 & 0 & 0 & 0 \\
   I & 0 & 0 & 0 & 0 & 0 & 0 & 0 \\
   I & 0 & 0 & 0 & 0 & 0 & 0 & 0 \\
   I & 0 & 0 & 0 & 0 & 0 & 0 & 0 \\
   I & 0 & 0 & 0 & 0 & 0 & 0 & 0 \\
   I & 0 & 0 & 0 & 0 & 0 & 0 & 0 \\
   0 & I & 0 & 0 & 0 & 0 & 0 & 0 \\
   0 & 0 & I & 0 & 0 & 0 & 0 & 0 \\
   0 & 0 & 0 & I & 0 & 0 & 0 & 0 \\
   0 & I & 0 & 0 & 0 & 0 & 0 & 0 \\
   0 & I & 0 & 0 & 0 & 0 & 0 & 0 \\
   0 & I & 0 & 0 & 0 & 0 & 0 & 0 \\
   0 & 0 & I & 0 & 0 & 0 & 0 & 0 \\
   0 & 0 & I & 0 & 0 & 0 & 0 & 0 \\
   0 & 0 & I & 0 & 0 & 0 & 0 & 0 \\
   0 & 0 & 0 & I & 0 & 0 & 0 & 0 \\
   0 & 0 & 0 & I & 0 & 0 & 0 & 0 \\
   0 & 0 & 0 & I & 0 & 0 & 0 & 0 \\
   0 & 0 & 0 & 0 & I & 0 & 0 & 0 \\
   0 & 0 & 0 & 0 & 0 & I & 0 & 0 \\
   0 & 0 & 0 & 0 & 0 & 0 & I & 0 \\
   0 & 0 & 0 & 0 & I & 0 & 0 & 0 \\
   0 & 0 & 0 & 0 & 0 & I & 0 & 0 \\
   0 & 0 & 0 & 0 & 0 & 0 & I & 0 \\
   0 & 0 & 0 & 0 & 0 & 0 & 0 & I \\
\end{bmatrix}_{n\times(n-m)}
\begin{bmatrix}
   U_{C1} \\ U_{E1} \\ U_{E2} \\ U_{E3} \\ U_{F1} \\ U_{F2} \\ U_{F3} \\ U_{int}
\end{bmatrix}_{(n-m)\times1},
\ee
\begin{figure}
    \centering
    \includegraphics[width=\linewidth]{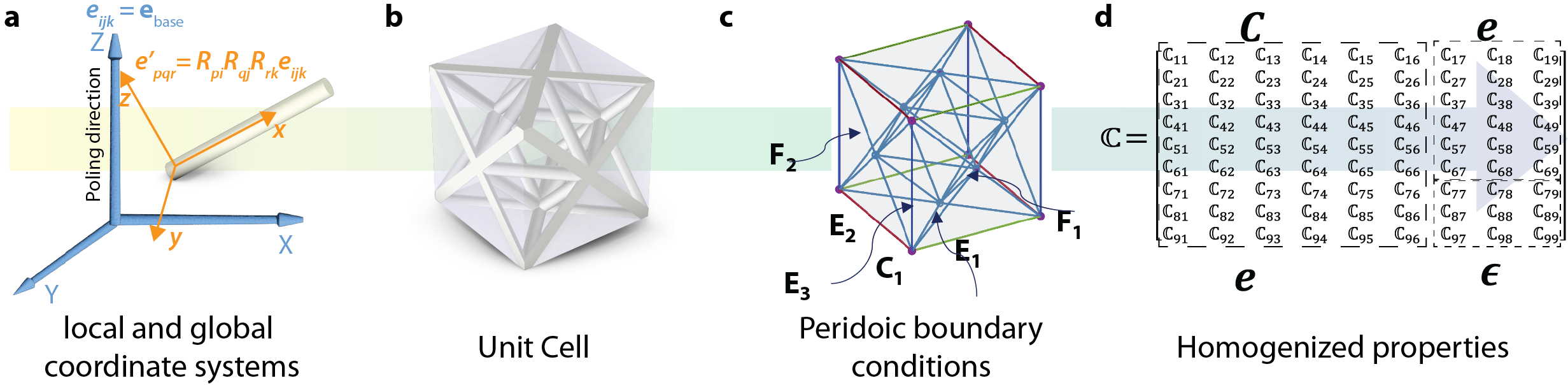}
    \captionsetup{justification=justified}
    \caption{\textbf{Schematic representation of the homogenization process.} \textbf{a-b.} The unit cell is first discretized using 1D Bernoulli-Euler elements. \textbf{c.} The exterior nodes are categorized into corner, face, and edge nodes, $C_1, E_1, E_2, E_3,$ and $F_1, F_2, F_3$ being the reference corner, edges, and faces. \textbf{d.} Finally, using the asymptotic homogenization approach, the homogenized electromechanical property tensor is computed.}
    \label{fig:PBC}
\end{figure}
where $n$ is the total number of dofs, m is the number of master dofs and $(n-m)$ is the number of coupled dofs. $U_{C_1} - U_{C_8}$ are the dofs of the nodes at corners $C_1-C_8$, $U_{E_1}-U_{E_{12}}$ are the dofs of the nodes at the edges $E_1-E_{12}$, $U_{F_1}-U_{F_6}$ are the dofs of the nodes at faces $F_1-F_6$, and $U_{int}$ are the dofs of rest of the nodes, i.e., interior nodes. PBCs are applied by coupling all corner nodes to node $C_1$, parallel edges to their corresponding reference edge among $E_1$, $E_2$, and $E_3$, and the opposing faces to the reference faces among $F_1$, $F_2$, and $F_3$. The reference corners, edges, and faces are as shown in Figure \ref{fig:PBC}.

Let $\dsC$ denote the matrix on the right side of Eq. 3 of the main article, which contains homogenized elastic, piezoelectric, and electrical stiffness coefficients. Its components can be computed as
\be 
\label{eq:CeffSI}
\dsC_{ijkl}=\frac{1}{V}(\bfchi^{0(ij)} - \Tilde{\bfchi}^{ij})^T (\bff^{kl}-\bff^{*kl}),
\ee
where $V$ is the volume of the unit cell, $\chi$, and $\bff$ are the nodal displacements and internal force vectors. $\bfchi^{0(ij)}$ are the initial displacements/electric potential, equivalent to unit strain/electric field applied to the nodes. To obtain the effective properties, three solution steps are carried out in the following sequence: 1) $\bfchi^{0(ij)}$ are applied to all the nodes in the unit cell and the generated internal nodal forces are computed as $\bff^{kl}=\bfK\bfchi^{0(ij)}$, 2) this nodal force vector is applied under PBCs to compute $\bfchi^{ij}=\Tilde{\bfK}^{-1}\Tilde{\bff}^{kl}$, where $\Tilde{\bfK}=\bfT^T\bfK\bfT$, and finally, 3) the computed periodic displacement vector is applied to the system without periodic boundary conditions to compute $\bff^{*(kl)}=\bfK\Tilde{X}^{ij}$. Effective coefficients are then computed by replacing these vectors into eq. \eqref{eq:CeffSI}. The FE homogenization gives the piezoelectric tensor in the stress charge form, which can be used to compute strain charge coefficients, $\bfd$ as
\be
\label{eq:etod}
 \bfd=\bfe.\bfS ,
\ee
where $\bfS = \bfC^{-1}$ is the compliance matrix in Voigt notation.

\section{ML-based design optimization}
\subsection{Design space I}
The machine learning models and optimization procedures were trained using specific hyperparameters as outlined in Table \ref{tab:parameters1}.
The optimization process begins by selecting the $S$ best initial guesses from the dataset as starting points.  These input features are concatenated and passed through the neural network to predict the corresponding piezoelectric response. To compute the objective, the predicted piezoelectric matrix is unnormalized to align with the correct physical scales. 

The optimization framework employs Gumbel softmax to convert categorical variables—such as lattice types and the number of tessellations—into discrete one-hot encoded representations. Throughout the optimization, the feature set is iteratively updated using the Adam optimizer. The stretch-limiting constraints for the design parameters $\bfU_I$ and $\bfV_I$, are applied as a pre-processing step at each iteration before the property prediction and gradient-based optimization step.

\begin{table}[ht]
\centering
\caption{List of parameters used for the first ML frameworks' training and design optimization protocols.} 
\label{tab:parameters1}
\begin{tabular}{lcc}
\hline
\multicolumn{1}{l}{\textbf{Parameter}} & \textbf{Notation} & \textbf{Value} \\ \hline
\textit{ML framework \#1 hyperparameters:}\\
$\quad$ Feature scaling &$-$ &  none, min-max-scaling \\
$\quad$ Input dimension forward NN &$-$ &  $46$ \\
$\quad$ Hidden dimensions &$-$ &  $1024,1024,1024,1024,1024,1024$ \\
$\quad$ Activation function hidden layer & $-$ & Leaky ReLU \\
$\quad$ Output dimension & $-$ &$18$ \\
$\quad$ Learning rate &$-$ & $0.001$ \\
$\quad$ Batch size &$-$ & $8096$ \\
$\quad$ Number of epochs &$-$ & $300$ \\
$\quad$ Optimizer &$-$ & Adam\cite{kingma2014adam}\\
$\quad$ Train/Test split & $-$ & $0.99/0.01$ \\\hline
\textit{Optimization setup hyperparameters:}\\
$\quad$ Magnitude regularization parameter & $\lambda_1$    & $0.3$\\
$\quad$ Shrinkage regularization parameter & $\lambda_2$    & $0.3$\\
$\quad$ Stretch scaling & $\alpha$    & $0.5$\\
$\quad$ Number of guesses & $S$    & $150$\\
$\quad$ Number of epochs & $E$    & $10000$  \\
$\quad$ Optimizer learning rate & $-$ & $0.01$ \\
$\quad$ Optimizer &$-$ & Adam\cite{kingma2014adam}\\
\end{tabular}
\end{table}

\subsection{Design space II}
The specific hyperparameters of the second ML framework and the optimization framework are outlined in Table \ref{tab:parameters2}.
The optimization process involves adjusting the latent variable $\bfz$ to meet specific performance criteria. Initially, we choose a set of S initial designs from the training dataset that perform the best for the objective we aim to optimize. Then, the structural data is encoded into the latent space using the encoder. Since the encoding process is stochastic we use Q different random seeds to encode the initial guesses into the latent space. The resulting latent representations are used as the starting points of the optimization.

\begin{figure}[ht]
    \centering
    \includegraphics[width=\linewidth, height=\textheight, keepaspectratio]{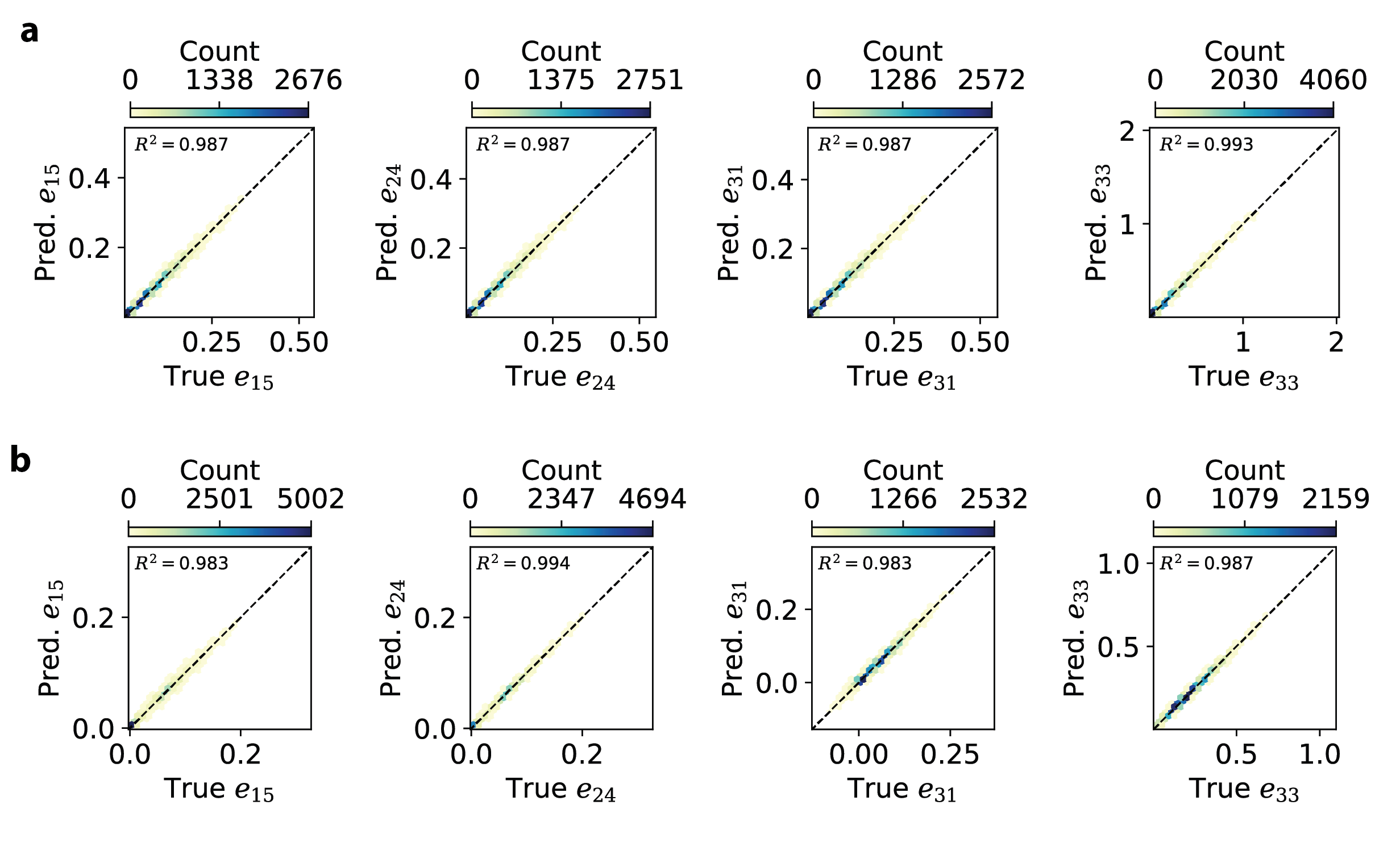}
    \captionsetup{justification=justified}
    \caption{\textbf{The parity plots for property prediction of datasets a. I and b. II.} Here, any point lying on the unit slope line means a perfect match between predicted and true values.}
    \label{fig:Trace}
\end{figure}

At each step, we decode the current latent representation and re-encode it back, analogously to the original optimization procedure\cite{zheng2023unifying}. From the re-encoded structure, we predict the piezoelectric matrix components and compute the objective. The optimization process adjusts the latent variables using the Adam optimizer\cite{kingma2014adam}, seeking to enhance the designs' performance relative to the specified criteria. Figures \ref{fig:Trace}\textbf{a} and \textbf{b} show the parity plots for the property predictors of dataset I and II, respectively.

\begin{table}[ht]
\centering
\caption{List of parameters used for the second ML frameworks' training and design optimization protocols.} 
\label{tab:parameters2}
\begin{tabular}{lcc}
\hline
\multicolumn{1}{l}{\textbf{Parameter}} & \textbf{Notation} & \textbf{Value} \\ \hline
\textit{ML framework \#2 hyperparameters:}\\
$\quad$ Feature scaling &$-$ &  none, min-max-scaling \\
$\quad$ Connectivity encoder input dimension &$-$ &  $278$ \\
$\quad$ Node position encoder input dimension &$-$ &  $27$ \\
$\quad$ Connectivity encoder hidden dimensions &$-$ &  $512,512,512,128$ \\
$\quad$ Node position encoder hidden dimensions &$-$ &  $640,640,640,512$ \\
$\quad$ Connectivity decoder input dimension &$-$ &  $32+8$ \\
$\quad$ Node position decoder input dimension &$-$ &  $32+8$ \\
$\quad$ Connectivity decoder hidden dimensions &$-$ &  $128,512,512,215$ \\
$\quad$ Node position decoder hidden dimensions &$-$ &  $256,512,640,640$ \\
$\quad$ Connectivity decoder output dimensions &$-$ &  $278$ \\
$\quad$ Node position decoder output dimensions &$-$ &  $27$ \\
$\quad$ Property predictor hidden dimensions &$-$ &  $400,800,1000,400,400,200$ \\
$\quad$ Property Predictor output dimensions &$-$ &  $5$ \\
$\quad$ Activation function hidden layer &$-$ &  ReLU \\
$\quad$ Connectivity (marg.) latent dimension & $-$ &$8$ \\
$\quad$ Node position (marg.) latent dimension & $-$ &$8$ \\
$\quad$ Overlapping latent dimension & $-$ &$32$ \\
$\quad$ Total latent dimension & $-$ &$48$ \\
$\quad$ Learning rate &$-$ & $0.0005$ \\
$\quad$ Batch size &$-$ & $8$ \\
$\quad$ Number of epochs &$-$ & $200$ \\
$\quad$ Optimizer &$-$ & Adam\cite{kingma2014adam}\\
$\quad$ Train/Test split & $-$ & $0.97/0.03$ \\\hline
\textit{Optimization setup hyperparameters:}\\
$\quad$ Number of guesses & $S$    & $1$\\
$\quad$ Number of random seeds for initial guess & $Q$    & $150$  \\
$\quad$ Number of epochs & $E$    & $30000$  \\
$\quad$ Optimizer learning rate & $-$ & $0.0001$ \\
$\quad$ Optimizer &$-$ & Adam\cite{kingma2014adam}\\
\end{tabular}
\end{table}

\section{Dataset exploration}

\begin{figure}[ht]
    \centering
    \includegraphics[width=\linewidth, height=\textheight, keepaspectratio]{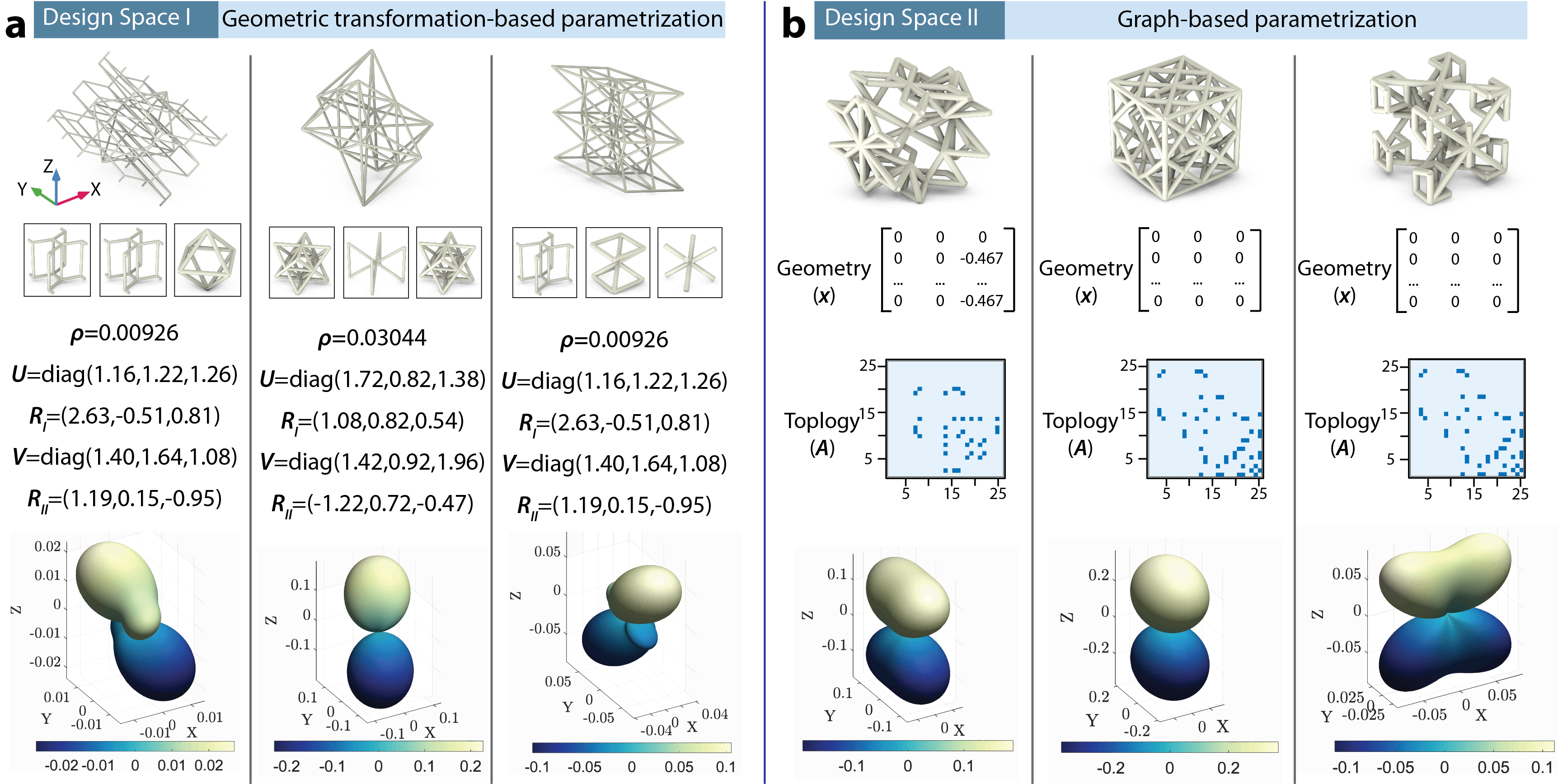}
    \captionsetup{justification=justified}
    \caption{\textbf{Process for generating dataset from design space I and II.} \textbf{a.} Design space I based on the overlapping and affine transformations of seven primitive unit cells. Three example unit cells and their design parameters are shown along with their corresponding $e_{33}$ surfaces. \textbf{b.} Design space II, based on the graph representation of unit cells. Three example unit cells are shown along with their geometric (in terms of the 27 nodal perturbations) and topological (connectivity matrix) parameters and corresponding $e_{33}$ surfaces.}
    \label{fig:Dataset_generation}
\end{figure}

Both the datasets used in this study consist of a wide range of unit cells with different anisotropies of piezoelectric tensor $\bfe$. While dataset I provides the full non-zero piezoelectric tensor, dataset II provides a broader range of topologies within the constraints of a cubic symmetry and hence only five original non-zero piezoelectric coefficients. \figureautorefname~\ref{fig:Dataset_generation}\textbf{a} and \textbf{b} show the unit cell generation process from a given set of parameters for dataset I and dataset II, respectively.

\begin{figure}[ht]
\vspace{-32 pt}
    \centering
    \includegraphics[width=\linewidth, height=\textheight, keepaspectratio]{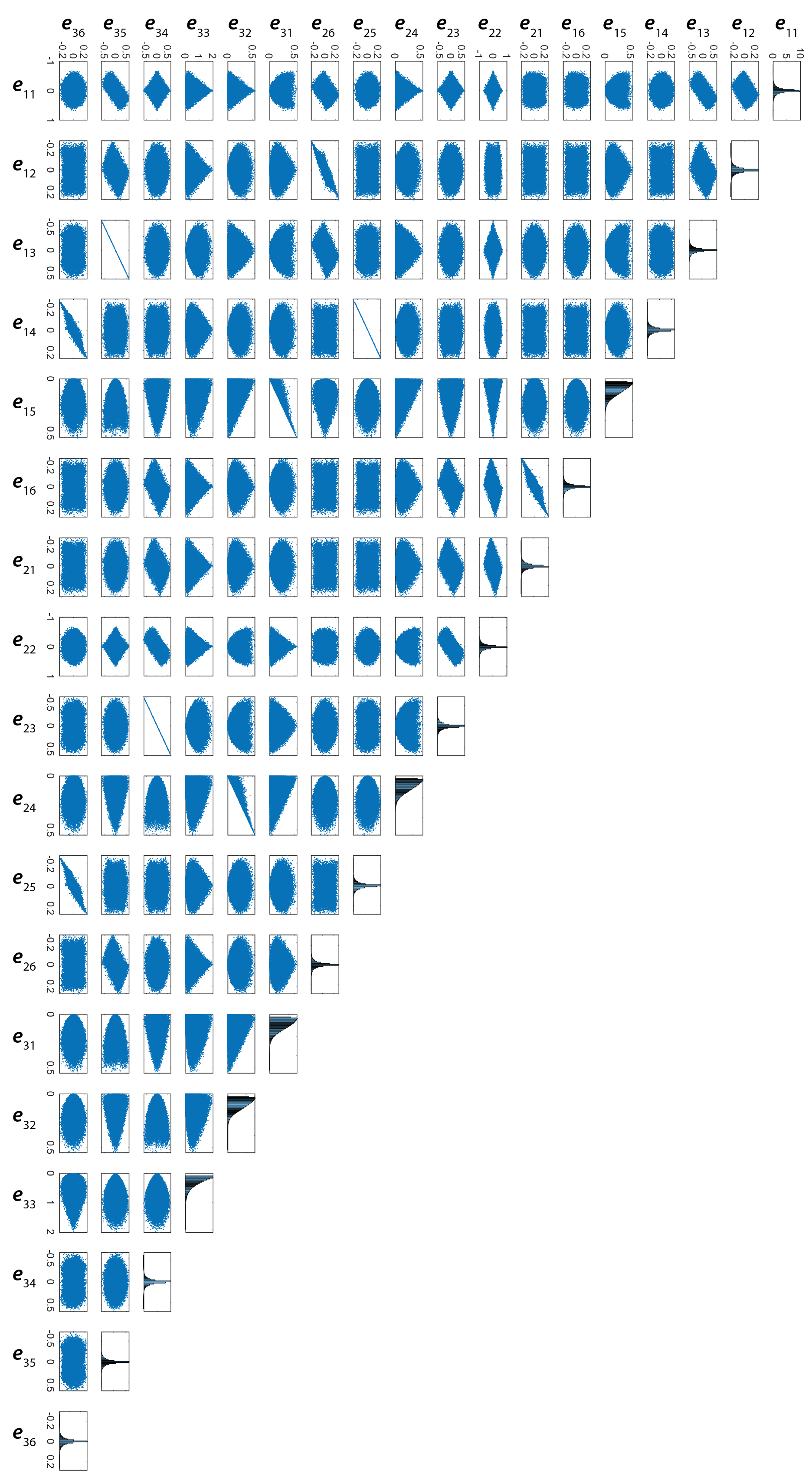}
    \captionsetup{justification=justified}
    \caption{\textbf{Pair plots of the full piezoelectric tensor $\bfe$ for dataset I.} Each plot represents the distribution of all the unit cells in the 2D space of piezoelectric coefficients on its horizontal and vertical axes. A concentration of points along the diagonal, such as $e_{13}$ vs $e_{15}$, of a plot represents a high correlation between the coefficients pertaining to the plot.}
    \label{fig:DS1_e}
\end{figure}

\figureautorefname~\ref{fig:DS1_e} shows the pair-plots of all 18 coefficients in the dataset I. While a vast landscape of property combinations is covered, some coefficients show strong correlations, e.g., $e_{25}-e_{14}$, $e_{35}-e_{13}$, and $e_{15}-e_{31}$. This is due to the fact that although a vast variety of geometries are generated, they originate from geometric transformations of 262 unique topologies. Since the piezoelectric properties are influenced by both the geometry and topologies of the unit cells, some correlation existing in the elementary topologies can influence the correlation in the whole dataset. 

\begin{figure}[ht]
    \centering
    \includegraphics[width=\linewidth, height=\textheight, keepaspectratio]{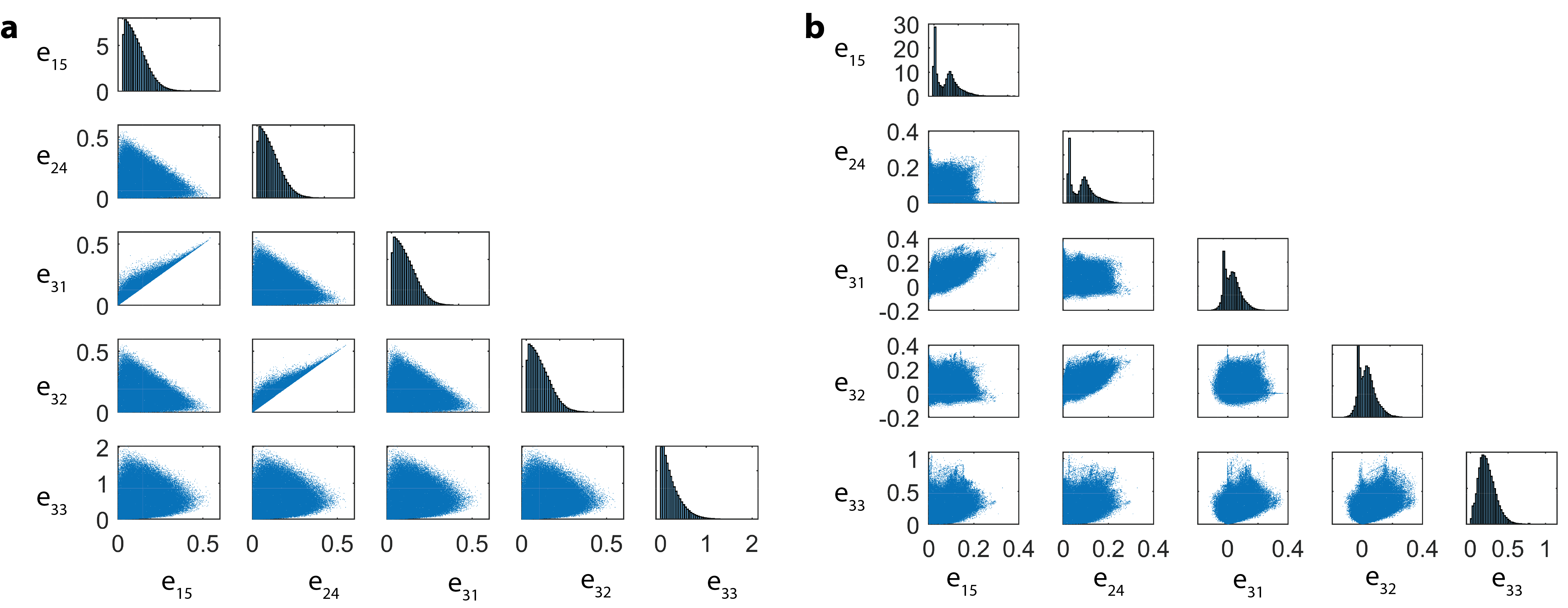}
    \captionsetup{justification=justified}
    \caption{\textbf{Pair plots of the five piezoelectric coefficients for} \textbf{a.} dataset I, and \textbf{b.} dataset II.}
    \label{fig:DS2_e}
\end{figure}

As mentioned before, dataset II is based on cubic unit cell volumes and thus only has five non-zero piezoelectric coefficients as the base material. However, due to the graph representation, a larger variety of topologies is obtained. This leads to a broader range of values of piezoelectric coefficients. \figureautorefname~\ref{fig:DS2_e} shows a comparison of the five piezoelectric coefficients of both datasets. As can be seen from the pair plots, dataset I shows correlations between different coefficients with approximately triangular patterns in the distribution of lattices. On the other hand, dataset II has more arbitrary distributions of the datapoints with no obvious correlations between different coefficients. Moreover, while dataset I only shows positive values of all five coefficients, dataset II exhibits both positive and negative values of $e_{31}$ and $e_{32}$.

\section{Additional examples}
\begin{figure}[ht]
    \centering
    \includegraphics[width=\linewidth, height=\textheight, keepaspectratio]{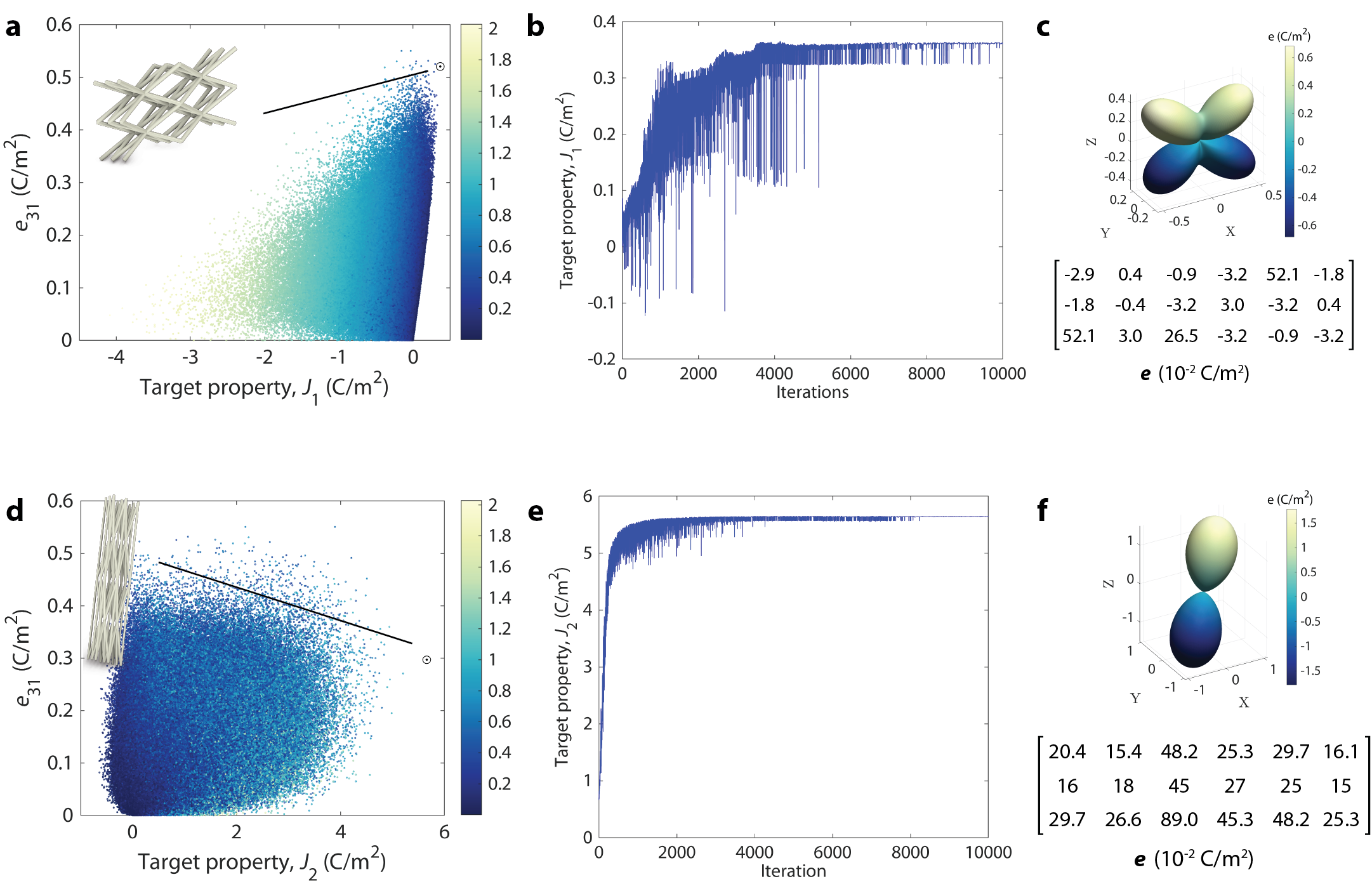}
    \captionsetup{justification=justified}
    % \caption{\textbf{Additional examples of tailored piezoelectric responses achieved with the ML-based optimization.} Unit cells with \textbf{A.} maximized $J_1$ for unidirectional piezoelectricity in $J_1$-$e_{31}$, landscape. The unit cell lies outside the property space of the training dataset.\textbf{B.} The trace followed by the optimization algorithm for achieving the optimized design for $J_1$ and \textbf{C} the piezoelectric surface (top) and the piezoelectric matrix (bottom). \textbf{D} Example of maximized $J_2$ for omnidirectional or fully anisotropic piezoelectricity, showing the optimized unit cell in $J_2$-$e_31$ landscape. Similar to the first example, the optimized unit cell lies well beyond the training dataset. \textbf{E}, and \textbf{F} show the optimization trace and the piezoelectric response of this optimized unit cell.}
    \caption{\textbf{Additional examples of tailored piezoelectric responses achieved with the ML-based optimization.} \textbf{a.} Unit cells with maximized $J_1$ for unidirectional piezoelectricity in the $J_1$ vs. $e_{31}$ landscape. The unit cell lies outside the property space of the training dataset. \textbf{b.} The optimization trajectory to achieve the optimized design for $J_1$. \textbf{c.} The piezoelectric surface (top) and the piezoelectric matrix (bottom). \textbf{d.} Example of maximized $J_2$ for omnidirectional or fully anisotropic piezoelectricity, showing the optimized unit cell in the $J_2$ vs. $e_{31}$ landscape. Similar to the first example, the optimized unit cell lies well beyond the training dataset. \textbf{e.} The optimization trajectory for achieving the optimized design for $J_2$. \textbf{f.} The piezoelectric response of this optimized unit cell.}
    \label{fig:Additional_Examples}
\end{figure}

In addition to the optimization examples shown in the main article, \figureautorefname~\ref{fig:Additional_Examples}~shows additional unit cells with exotic properties achieved through ML-based optimization. In the first example (\figureautorefname~\ref{fig:Additional_Examples}\textbf{a-c}), we show a lattice optimized for achieving unidirectional response in response to an applied electric field in direction-3. This is achieved by solving the following optimization problem:
\begin{align}
\begin{split}
\label{eq:Example-A}
\Theta^* \leftarrow &\ \argmaxA_\Theta\ J_1, \qquad \text{with}\quad \bfe = \calF_\omega(\Theta) \quad \text{and} \\
&\ J_1 =\left(e_{31}-e_{32}^2-e_{33}^2 \right) - \lambda_2 \left(\sum_{\substack{i=1}}^{6} e_{1i}^2 + \sum_{i=1}^{6} e_{2i}^2 + \sum_{i=4}^{6} e_{3i}^2\right)
\end{split}
\end{align}

In the second example, as shown in  \figureautorefname~\ref{fig:Additional_Examples}\textbf{d-f}, the unit cell design is optimized to achieve omnidirectional piezoelectricity or full anisotropy of the piezoelectric behavior by solving:

\be
\begin{split}
\label{eq:Example-B}
\Theta^* \leftarrow  \argmaxA_\Theta\ J_2, \qquad \text{with}\quad \bfe = \calF_\omega(\Theta) \quad \text{and} \quad
 J_2 =\left(\sum_{i=1}^{3}\sum_{j=1}^6 e_{ij} \right).
\end{split}
\ee
\figureautorefname~\ref{fig:Additional_Examples}\textbf{b} and \textbf{e} show the traces followed by the optimization algorithm for $J_1$ and $J_2$, respectively. \figureautorefname~\ref{fig:Additional_Examples}\textbf{c} and \textbf{f} show the piezoelectric surfaces (top) and the piezoelectric matrices (bottom) for these examples.

\section{Principal component analysis (PCA)}
\begin{figure}
    \centering
    \includegraphics[width=\linewidth, height=\textheight, keepaspectratio]{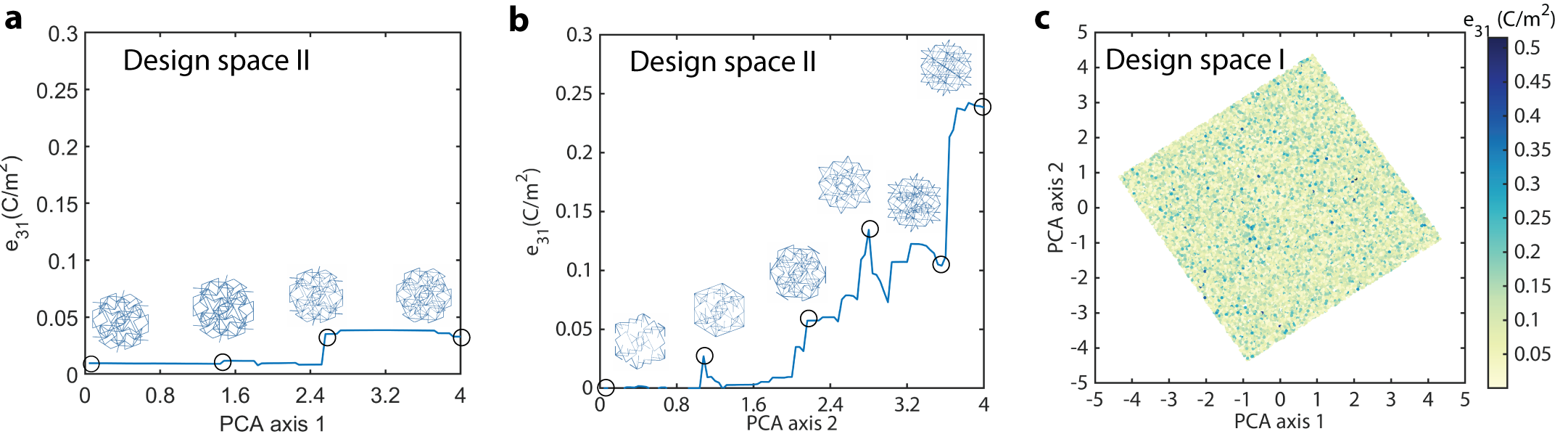}
    \captionsetup{justification=justified}
    \caption{\textbf{Principal component analysis of the two design spaces.} For the design space II, \textbf{a.} shows the variation of $e_{31}$ along the principal component axis 1, showing small variation in $e_31$, \textbf{b} variation of $e_{31}$ along the principal component axis 2, with significant variations in the unit cell design as well as $e_{31}$. \textbf{c.} PCA of design space I in the parameter space. Due to the abrupt changes in piezoelectric properties with design parameters, variation of $e_{31}$ does not exhibit a discernible trend.}
    \label{fig:PCA_SI}
\end{figure}
\clearpage

\bibliography{Bib}